\documentclass[rmp,preprint]{revtex4}
\usepackage{layout}

\usepackage{amsmath}
\usepackage{amsfonts}
\usepackage{amssymb}
\usepackage{graphicx}
\usepackage{epsfig}

\begin{document}

\title{Reversals of a large scale field generated over a turbulent background}

\author{B. Gallet, J. Herault, C. Laroche, F. P\'etr\'elis, S. Fauve\\
\vspace{6pt} Laboratoire de Physique Statistique, Ecole Normale Sup\'erieure,  CNRS,  24 rue  Lhomond, 75005 Paris France}

\begin{abstract}
We present a study of several systems in which a large scale field is generated over a turbulent background. 
These large scale fields usually break a symmetry of the forcing by selecting a direction. 
Under certain conditions,  the large scale field displays reversals so that  the symmetry of the forcing is recovered statistically. 
We present examples of such dynamics in the context of the dynamo instability, of two dimensional turbulent Kolmogorov flows and of turbulent Rayleigh-B\'enard convection. In these systems reversals  occur respectively for the dynamo magnetic field, for  the large scale circulation generated by a periodic forcing in space and for the large scale roll generated by turbulent thermal convection. We  compare the mechanisms involved and  show that their properties depend on some symmetries of the system and on the way they are  broken. 
\end{abstract}

\maketitle

\section{Introduction}

It is often believed that fully developed turbulent flows are statistically invariant under symmetry transformations of the forcing that generates them. Indeed, although the transition to turbulence involves successive bifurcations, each of them breaking some spatial or temporal symmetry, it is often observed that these symmetries are recovered in a statistical sense when the Reynolds number is increased further. The common belief is that strong turbulent fluctuations trigger random transitions between symmetric solutions and thus prevent the flow from staying in a regime with a broken symmetry. In other words, a strongly turbulent flow is expected to explore all the available phase space. However, it has been known since a long time that this is not so simple and that clear-cut transitions keep occurring within the strong turbulent regime. The oldest example is provided by the drag crisis (see for instance Tritton 1977). The mean drag of a sphere or a cylinder in a turbulent flow drops suddenly for a critical value of the Reynolds number $Re$ of order $10^5$ (the pre-factor depending on surface properties). This corresponds to a transition where the mean flow pattern changes, the wake becoming narrower. Another example is related to Rayleigh-B\'enard convection, i.e. the flow generated by heating from below a horizontal layer of fluid. It has been observed that for a Rayleigh number of order $10^6$, i.e. roughly $1000$ times larger than the critical Rayleigh number for the onset of convection, a large scale flow is generated with a horizontal extension equal to the length of the container (Krishnamurti \& Howard 1981). Von Karman swirling flows, i.e. flows generated in a cylindrical volume by the rotation of two co-axial disks, also display turbulent bifurcations. In the case of co-rotating disks, an axisymmetric mean flow with a strong axial vortex is observed. When the rotation rates are varied, this flow breaks axisymmetry, thus generating a roughly periodic modulation of the turbulent velocity field superimposed to turbulent fluctuations with $Re \sim 10^5$ (Labb\'e \textit{et al.} 1996). In the case of disks counter-rotating at the same frequency, the forcing is symmetric with respect to a rotation of angle $\pi$ about any radial axis in the mid-plane between the two disks (see below). It has been found that this symmetry can be broken through a bifurcation that occurs for Reynolds number in the range $10^5$ to $10^6$ (Ravelet \textit{et al.} 2004).

There are a lot of other examples of transitions leading to broken symmetries in strongly  turbulent flows. Bifurcations from a turbulent regime are therefore commonly observed. 
However, in contrast to bifurcations from stationary or space and time periodic flows, that  are well documented and for which well-known techniques exist to handle both the linear stability problem and the weakly nonlinear bifurcated regime, the concept of bifurcation from a fully developed turbulent flow is more questionable even at the level of a clear cut definition. In this respect, the dynamo effect in a liquid metal provides a very interesting example in which a magnetic field is generated by an instability process that occurs when the kinetic Reynolds number of the flow is usually larger than $10^6$. Although the experiments involve some cost and technical difficulties, once the dynamo regime is reached, the dynamics of the magnetic field can be easily measured. Then, several interesting problems can be addressed. One of them is related to the possibility of anomalous scalings of the magnetic energy density above the dynamo threshold due to turbulent fluctuations (P\'etr\'elis \textit{et al.} 2007). Another one concerns the effect of turbulent fluctuations on the dynamics of large scale magnetic field. The symmetry ${\bf B}({\bf r}, t) \rightarrow - {\bf B}({\bf r}, t)$ of the equations of magnetohydrodynamics is spontaneously broken at the dynamo onset. Slightly above the threshold of a supercritically bifurcating dynamo, the magnetic energy density is much smaller than the kinetic energy density of turbulent fluctuations. However, turbulent fluctuations are not able to trigger a direct transition from $B$ to $- B$ as would be observed in the naive analogy of a particle in a two-well potential in the presence of noise. In contrast, the VKS experiment has shown that the reversal trajectories display very robust features that are not smeared out by turbulent fluctuations (Berhanu \textit{et al.} 2007). In other words, the dynamics of the large scale magnetic field involves only a few modes that look weakly coupled to the turbulent background, i.e. the reversal dynamics takes place in a low dimensional phase space. Thus, the VKS dynamo provides an example in which a few magnetic modes are governed by a low dimensional dynamical system although they occur on a strongly turbulent background at kinetic Reynolds number larger than $10^6$.  

Similar features are also observed on purely hydrodynamical models and the same problems can be addressed. This is the case in many geophysical and astrophysical flows where an oscillatory or a weakly chaotic behaviour can occur in flows at huge Reynolds numbers. Some climatic phenomena indeed display a characteristic feature of low dimensional chaos: well defined patterns occur within a random temporal behaviour. Examples are atmospheric blockings that can affect the climate on a time scale of several days (Ghil \& Childress 1987) or El Nino events that occur every few years (Vallis 1986). The qualitative features of these phenomena have been often modelled using a few coupled variables such as mean temperature, wind or current. The quasi-biennial oscillation(QBO) provides a striking example of a large scale almost cyclic reversing flow in the otherwise  turbulent atmosphere. It is a roughly periodic oscillation in the strength and direction of the zonal (east-west) wind in the lower and middle stratosphere over the equator of the Earth's atmosphere.  It has been observed for more than 50 years in climatological records. Its period fluctuates, the mean being slightly larger than 2 years. A reversal first appears at an altitude of roughy $40$ km and then propagates downward at a rate of $1$ km/month. Thus, these reversals are related to a downward drifting pattern. The QBO has been understood to arise from the interaction between upward-propagating waves, generated in the troposphere, and the mean zonal flow at upper levels where the waves are dissipated (for a review, see Baldwin \textit{et al.}  2001). Reversals of large scale flows driven on a turbulent background, have been also observed in laboratory experiments, such as thermal convection (see Ahlers \textit{et al.} 2009 for a review) and Kolmogorov flows, i.e., quasi-two-dimensional flows generated by forcing an array of counter-rotating vortices (Sommeria 1986).
We first recall here old and recent observations related to bifurcations generating reversals of a magnetic field or of a mean flow in the presence of turbulent fluctuations and we discuss them in the framework presented above. Experimental results and models for reversals of the magnetic field are presented in section~\ref{dynamo}. Reversals of the large scale flow generated by Kolmogorov forcing are studied in section~\ref{Kolmogorov}. An experimental confirmation of Sommeria's results is provided and compared with a direct numerical simulation. A description of these reversals using a low dimensional dynamical system is given and their main features are compared to the ones observed for the magnetic field. Reversals of the large scale flow in Rayleigh-B\'enard convection are considered in section~\ref{convection}. After recalling the results obtained in the literature, we present a new experiment showing that three-dimensional perturbations can play an important role in the reversal process. Finally, we discuss the problem of reversals in relation to the phenomenon of drifting patterns and using the framework of amplitude equations. Thus, different types of bifurcations leading to reversals in this variety of systems can be easily identified. We conclude with a discussion on  the relative contributions of the deterministic dynamics and of the turbulent fluctuations to reversals observed in strongly turbulent flows.

\section{Reversals of the magnetic field generated in a fluid dynamo}\label{dynamo}

\subsection{Dynamos and reversals of the magnetic field}\label{reversals}

The dynamo instability is a generic mechanism that converts motion, {\it i.e.} kinetic energy into electric current, {\it i.e.} magnetic energy. This  process has been used for a long time to generate electricity from the motion of solid rotators. Larmor  (1919) proposed that in the  Sun a similar instability could be operating   that  would convert kinetic energy of  the flow of an electrically conducting fluid into magnetic energy. Nowadays  it is strongly believed that most astrophysical objects (planets, stars, galaxies, ...) generate a magnetic field  through a dynamo instability (Moffatt 1978).
 In the last decade, the first three experimental observations of fluid dynamos have been achieved  in turbulent flows of liquid sodium (Gailitis \textit{et al.}  2001, Stieglitz and Muller 2001, Monchaux \textit{et al.}  2007). 

Since the work of Brunhes (1906), it is known that the Earth magnetic field keeps a roughly  constant direction for long durations but from time to time it reverses and evolves toward the opposite direction. Reversals of the Earth
magnetic field have motivated a lot of studies ranging from paleomagnetism to numerical
simulations of models of the Earth interior (see for instance Dormy \textit{et al.} 2000, P\'etr\'elis and Fauve 2010).

\begin{figure}
\centering
\includegraphics[width=14cm,angle=0]{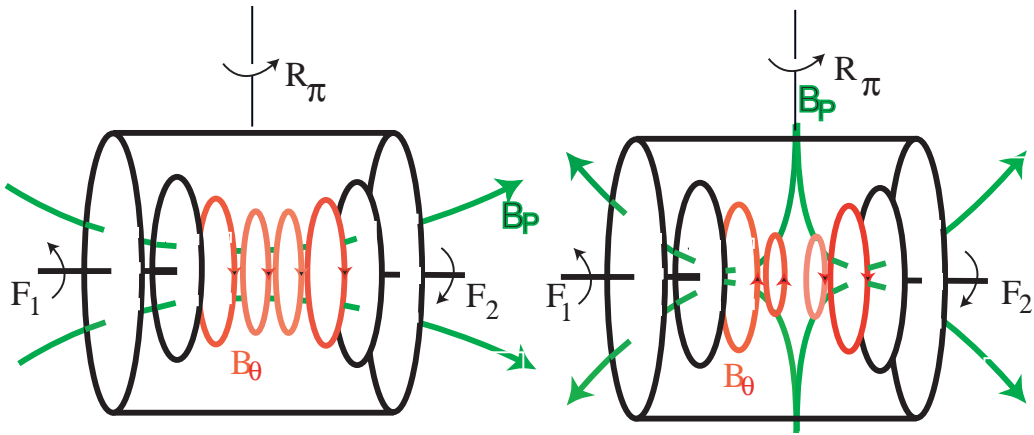}\\
\vspace{- 2cm}
\caption[]{Sketch of the large scales of the eigenmodes  of the VKS experiment. The two disks counter-rotate with frequency $F_1$ and $F_2$. Left: dipolar part of the magnetic mode. Right: quadrupolar part. Poloidal, $B_P$, and toroidal, $B_{\theta}$, components are sketched.}
\label{figmodevks}       
\end{figure}

\begin{figure}[htb!]
\centering
\includegraphics[width=10cm,height=7cm,angle=0]{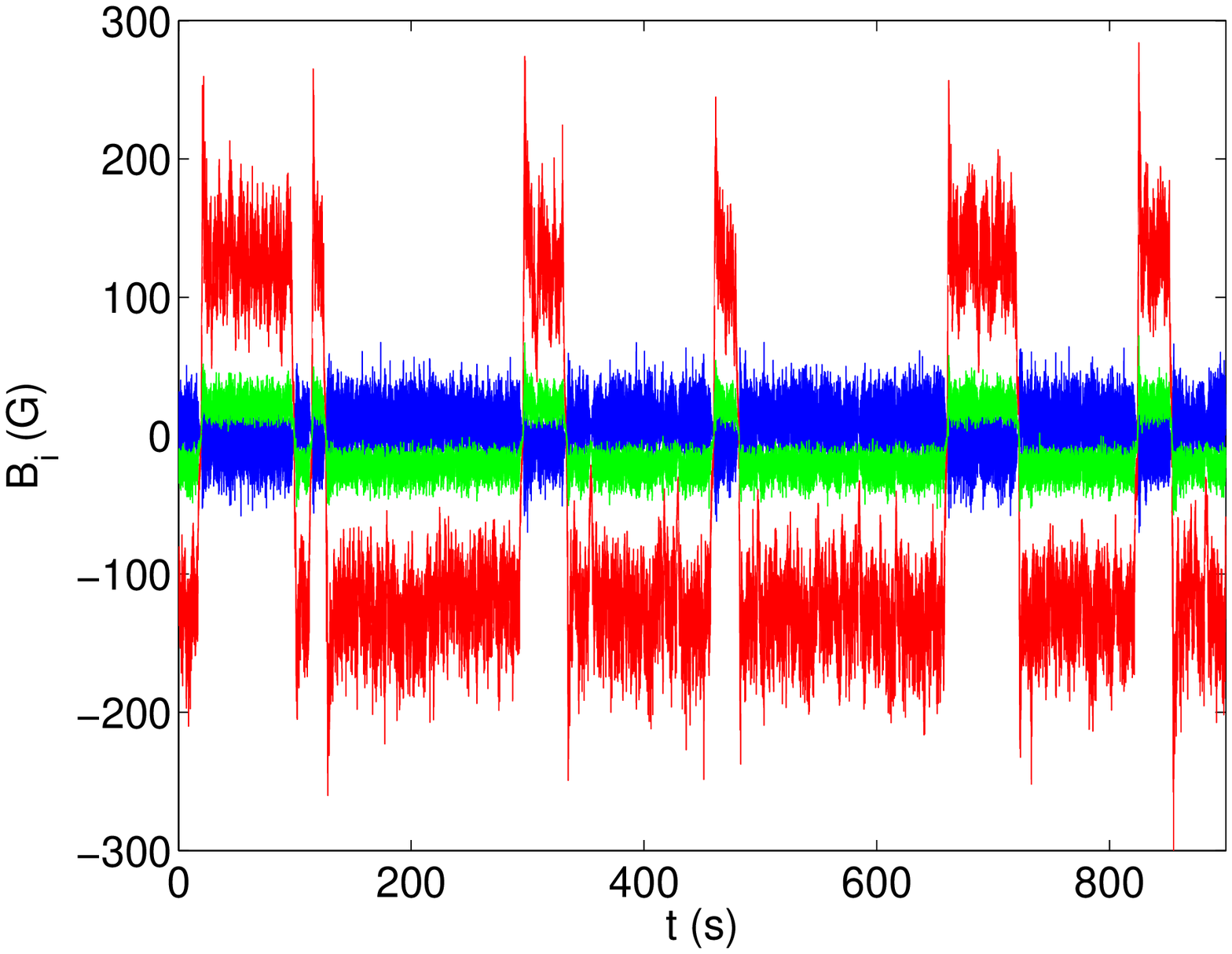}\includegraphics[width=6.5cm,height=6cm,angle=0]{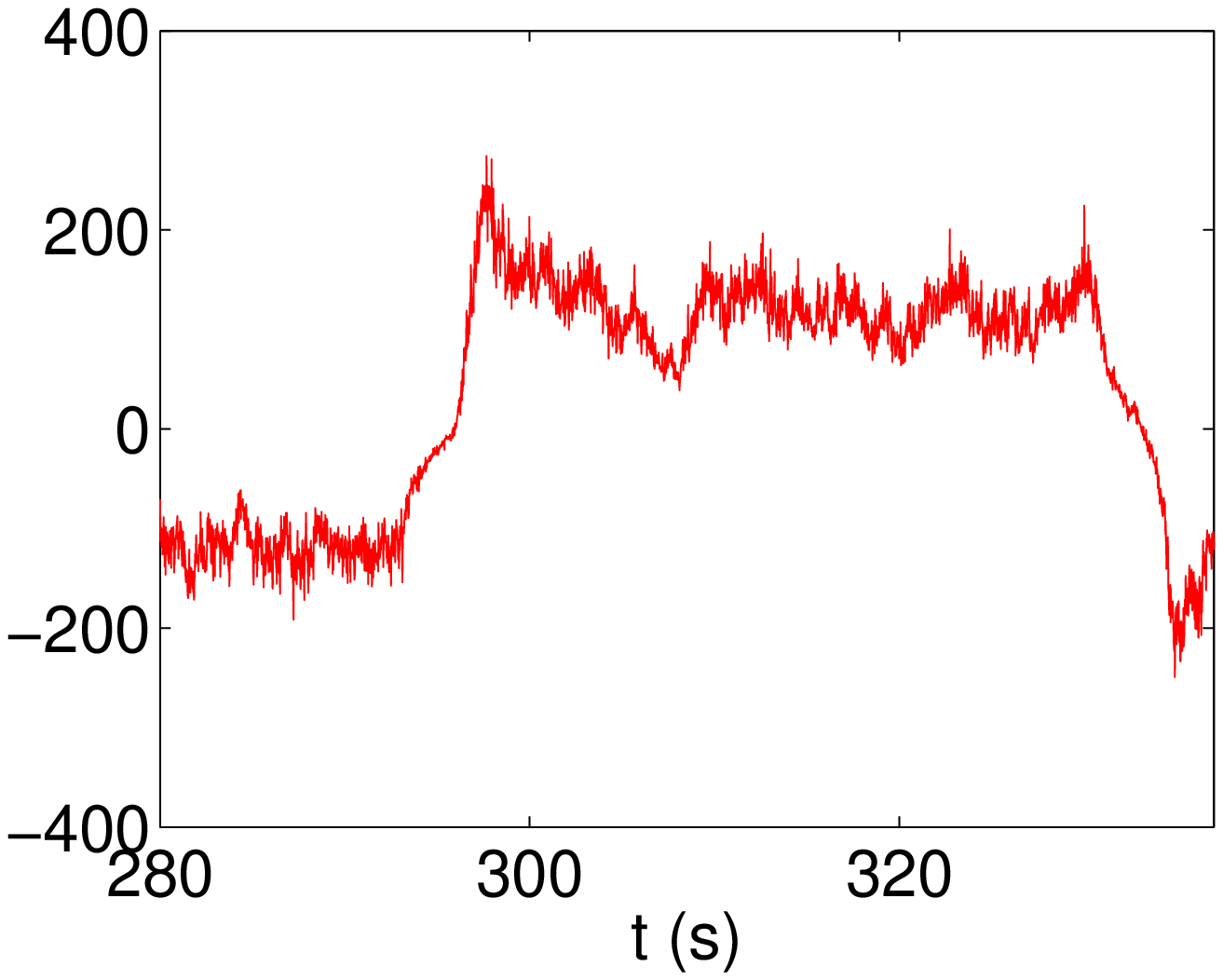}
\caption[]{Left: Time series of the three components of the magnetic field measured in the VKS experiment in the regime of reversals. The two propellers counterrotate with different speeds. Right: zoom on two reversals and an excursion (data from Berhanu \textit{et al.} 2007).}
\label{figvks}       
\end{figure}

Reversals of the magnetic field generated by a fluid dynamo have been observed only recently in laboratory experiments (the VKS experiment,  Berhanu \textit{et al.} 2007).
This experiment involves a turbulent swirling flow of liquid sodium, generated by  two soft iron impellers, counter-rotating at frequency $F_1$ (respectively $F_2$) in an inner copper cylinder, as sketched in fig. \ref{figmodevks}. When the disks counter-rotate with the same frequency $F$, a statistically stationary magnetic field is generated when $F$ is large enough. Its mean value involves a dominant poloidal dipolar component, ${\bf B}_P$, along the axis of rotation, together with a related azimuthal component $\bf{B}_{\theta}$, as displayed in fig. \ref{figmodevks} (left).  When the rotation frequencies are different, the magnetic field can display periodic or  random reversals. 	A time series of the field in the regime of  random reversals  is displayed in fig. \ref{figvks}.  The field fluctuates around some value for long durations, typically hundreds of seconds, and suddenly reverses within a few seconds. 
Although the kinetic Reynolds number is large, $Re \sim 5\, 10^6$ for these flows, it has been observed that the low dimensional dynamics of the magnetic field is not smeared out by strong turbulent fluctuations. In particular, all reversals involve the same transitional field morphology: the amplitude of the dipolar field first decreases. If it changes polarity, the amplitude increases on a faster time scale and then displays an overshoot before reaching its statistically stationary state. Otherwise, the magnetic field grows again with its direction unchanged.  The above features are also observed in recordings of the Earth's magnetic field, the aborted reversals being often called excursions (Valet \textit{et al.} 2005).

\subsection{A generic mechanism for reversals of the magnetic field}\label{model}

The most striking  feature of the VKS experiment is that time dependent magnetic fields are generated only when the impellers rotate at different frequencies (Berhanu \textit{et al.} 2007, Ravelet \textit{et al.}  2008). We have shown in P\'etr\'elis \& Fauve (2008) that this is related to the broken invariance under ${\cal R}_{\pi}$ when $F_1 \neq F_2$ (rotation of an angle $\pi$ along any radial axis in the mid-plane). In that case, symmetric and antisymmetric modes (under ${\cal R}_{\pi}$) are linearly coupled. Such modes are displayed in figure \ref{figmodevks}: a dipolar mode is changed to its opposite by ${\cal R}_{\pi}$, whereas a quadrupolar mode is unchanged.

Although the flow in the VKS experiment strongly differs from the one in the Earth's core, dipolar and quadrupolar modes can be defined in both cases (using different symmetries, see P\'etr\'elis {\it et al.} 2009). We  assume that the magnetic field is the sum of a dipolar component with an amplitude $D$ and a quadrupolar one, $Q$. We define $A=D+i\, Q$ and we assume that an expansion in power of  $A$  and its complex conjugate $\bar{A}$ is pertinent close to threshold in order to obtain an evolution equation for both modes. Taking into account the invariance ${\bf B} \rightarrow-{\bf B}$, {\it i.e.} $A\rightarrow -A$,  we obtain 
\begin{eqnarray}
\dot{A}&=&\mu A+\nu \bar{A}+\beta_1 A^3+\beta_2 A^2\bar{A}+\beta_3 A \bar{A}^2+\beta_4\bar{A}^3\,,
\label{eqdipquad}
\end{eqnarray}
where we limit the expansion to the lowest order nonlinearities.
In the general case, the coefficients are complex and depend on the experimental parameters.

Symmetry of the  experiment  with respect to ${\cal R}_{\pi}$ amounts to constraints on the coefficients. Applying the transformation ${\cal R}_{\pi}$ changes $D$  and  $Q$ in different ways: $D \rightarrow - D$, $Q \rightarrow  Q$, thus $A\rightarrow-\bar{A}$. We conclude that, in  the case of exact counter-rotation all the coefficients are real. More generally, the real parts are even and the imaginary parts are odd functions of the frequency difference $f=F_1-F_2$.

\begin{figure}
\centering
\includegraphics[width=10cm,angle=0]{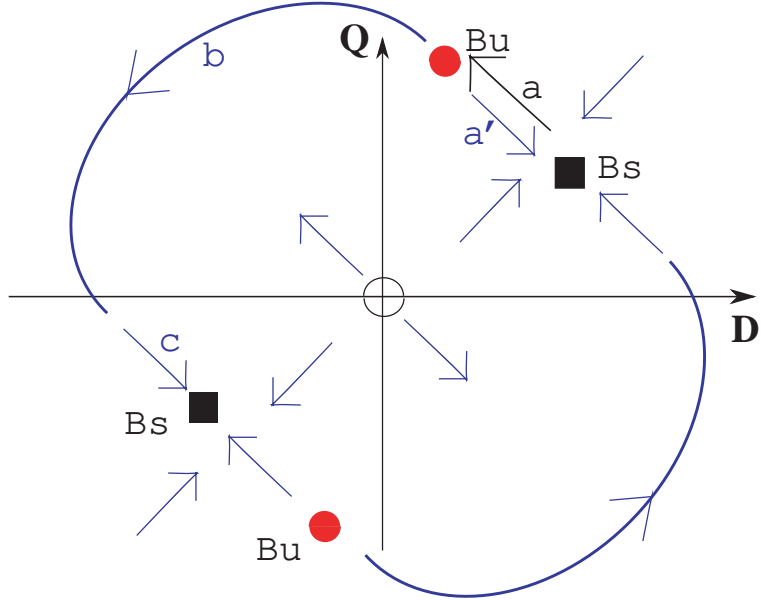}\\
\caption[]{A generic saddle-node bifurcation in a system with the ${\bf B} \rightarrow - {\bf B}$ invariance: below threshold, fluctuations can drive the system against its deterministic dynamics (phase a). If the effect of fluctuations is large enough, this generates a reversal (phases b and c). Otherwise, an excursion occurs (phase a').}
\label{saddlenode}       
\end{figure}

The coefficients of (\ref{eqdipquad}) can be chosen so that it has two stable dipolar solutions $\pm D$ and two unstable quadrupolar solutions $\pm Q$ when $f=0$. When $f$ is increased, these solutions become more and more mixed due to the increase in the strength of the coupling terms between dipolar and quadrupolar modes. Dipolar (respectively quadrupolar) solutions get a quadrupolar (respectively dipolar) component and give rise to the stable solutions $\pm B_s$ (respectively unstable solutions $\pm B_u$) displayed in figure \ref{saddlenode}.  
When $f$ is increased further, a saddle-node bifurcation can occur, i.e. the stable and unstable solutions collide by pairs and disappear (P\'etr\'elis \& Fauve 2008).  A  limit cycle is generated that connects the collision point, $B_c$, with its opposite.  


This provides an elementary mechanism for field reversals in the vicinity of a saddle-node bifurcation. First, in the absence of fluctuations, the limit cycle generated at the saddle-node bifurcation connects $\pm B_c$. This corresponds to periodic reversals. Slightly above the bifurcation threshold, the system spends most of the time close to the two states of opposite polarity $\pm B_c$. Second, in the presence of fluctuations, random reversals can be obtained slightly below the saddle-node bifurcation.  $B_u$ being very close to $B_s$, even a fluctuation of small intensity can drive the system to $B_u$ from which it can be attracted by $-B_s$, thus generating a reversal. 

\begin{figure}
\centering
\includegraphics[width=10cm,height=6cm,angle=0]{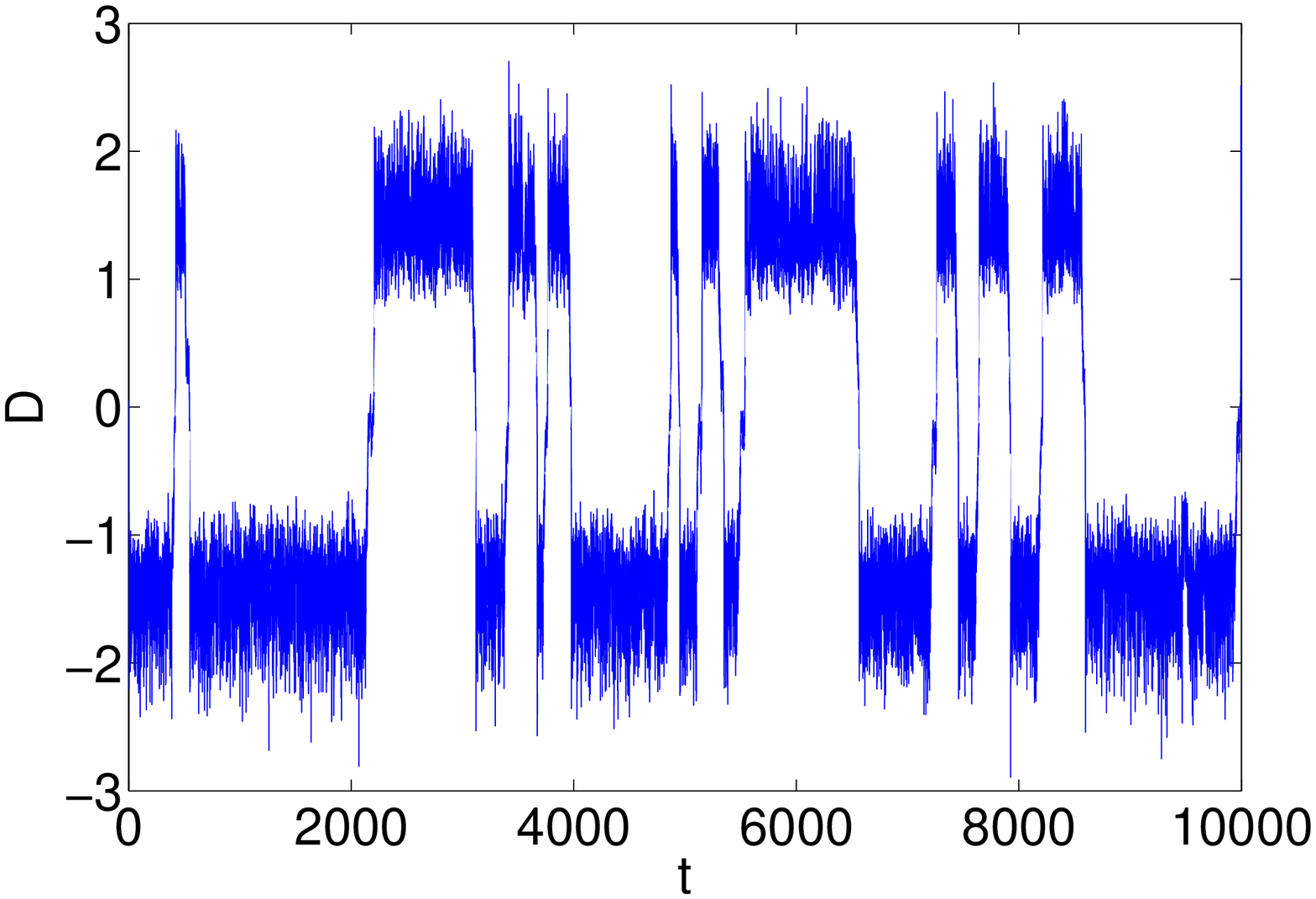}
\includegraphics[width=6cm,height=6cm,angle=0]{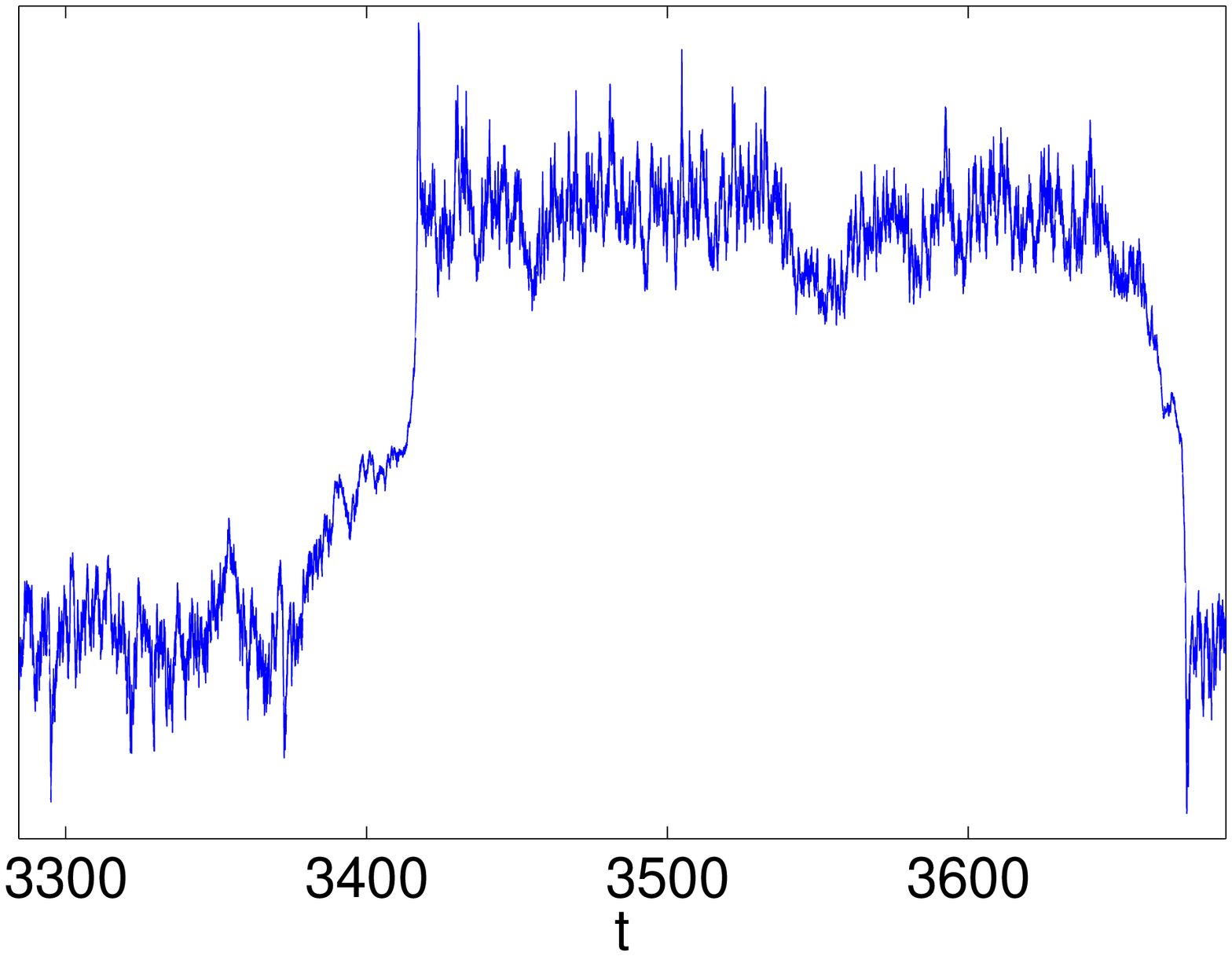}
\\
\caption[]{(left) Reversals of the magnetic field modelled by (\ref{eqdipquad}) with an extra noise term that takes into account the effect of the turbulent fluctuations. (right) Zoom on reversals and excursions.}
\label{reversalssim}       
\end{figure}

The effect of turbulent fluctuations on the dynamics of the two magnetic modes governed by (\ref{eqdipquad}) can be easily modelled by adding some noisy component to the coefficients. 
Random reversals are displayed in figure \ref{reversalssim} (left). The system spends most of the time close to the stable fixed points $\pm B_s$.
We observe in figure \ref{reversalssim}  (right) that a reversal consists of two phases. In the first phase, the system evolves from the stable point $B_s$ to the unstable point $B_u$ (in the phase space sketched in figure \ref{saddlenode}). The deterministic part of the dynamics acts against this evolution and the fluctuations are the motor of  the dynamics. This phase is thus slow. In the second phase, the system evolves from $B_u$ to $-B_s$, the deterministic part of the dynamics drives the system and this phase is faster. 

The behaviour of the system close to $B_s$ depends on the local flow in phase space. Close to the saddle-node bifurcation,  the position of $B_s$ and $B_u$  defines the slow direction of the dynamics. If a component of $B_u$ is smaller than the corresponding one of  $B_s$, that component displays an overshoot at the end of a reversal. In the opposite case, that component will increase at the beginning of a reversal. For instance, in the phase space sketched  in figure  \ref{saddlenode}, the component $D$ decreases at the end of a reversal and the signal displays an overshoot.  The component $Q$ increases just before a reversal.

For some fluctuations, the second phase does not connect $B_u$ to $-B_s$ but to $B_s$. It is an aborted reversal or an excursion in the context of the geodynamo.  Note that during the initial phase, a reversal and an excursion are identical. In the second phase, the approaches to the fixed point differ because the trajectory that links $B_u$ and $B_s$ is different form the trajectory that links $B_u$ and $-B_s$.  
In the case of  figure \ref{saddlenode}, the dipole  displays an overshoot at the end of  a reversal and  reaches smaller values during an excursion (see  figure \ref{reversalssim} right). In contrast the quadrupole displays an overshoot at the beginning of a reversal and reaches larger values during an excursion.  

The above mechanism explains several observations made in numerical simulations: a dipole-quadrupole interaction is clearly visible in the first reversals simulated by Glatzmaier \& Roberts (1995). They notice that ``the toroidal field is asymmetric with respect to the Equator before and after the reversal but is symmetric midway through the transition". The field has thus has a quadrupolar symmetry at the transition (see their figure 2). This has been confirmed by the simulations of Sarson \& Jones (1999) who find that reversals rely "heavily upon the interaction between dipole and quadrupole symmetries" and that they are triggered by the random emission of poleward light plumes, i.e. events that break the equatorial symmetry of the flow. Similar features have been observed by Wicht \& Olson (2004). Li  \textit{et al.} (2002) also emphasise that ``the dipole polarity can reverse only ... where the north-south symmetry of the convection pattern is broken" and that ``the quadrupole mode grows ... before the reversal". It has been also shown that if the flow or the magnetic field is forced to remain equatorially symmetric, then reversals do not occur (Nishikawa \&  Kusano 2008).  A direct numerical simulation of the dynamo generated by a flow driven in a sphere by two counter-rotating co-axial propellers (Gissinger \textit{et al.} 2010) has also displayed the following features. Reversals of the axial dipole occur only when the propellers are rotated at different rates. When the magnetic Prandtl number  ($P_m$) is small enough, they involve an axial quadrupole  and the dynamics of the dipolar and quadrupolar modes during a reversal is similar to the one observed in the VKS experiment. 

\subsection{A dynamical system displaying random reversals of the magnetic field}

The three modes that are important in the low  $P_m$  simulations are the dipole $D$, the quadrupole $Q$, and the zonal velocity mode $V$ that breaks the ${\cal R}_{\pi}$ symmetry. These modes transform as $D \rightarrow - D$, $Q \rightarrow Q$ and $V \rightarrow - V$ under the ${\cal R}_{\pi}$ symmetry. Keeping nonlinear terms up to quadratic order, we get the dynamical system (Gissinger \textit{et al.} 2010)
\begin{eqnarray}
\dot{D} &=& \mu D - V Q, \label{ampD}   \nonumber \\
\dot{Q} &=& - \nu Q + V D, \label{ampQ}  \nonumber \\
\dot{V} &=& \Gamma - V + Q D. \label{ampV}
\end{eqnarray}
\label{dqv}
A non zero value of $\Gamma$ is related to a forcing that breaks the ${\cal R}_{\pi}$ symmetry, i.e. propellers rotating at different speeds.

\begin{figure}[htb!]
\begin{center}
\includegraphics[width=9cm]{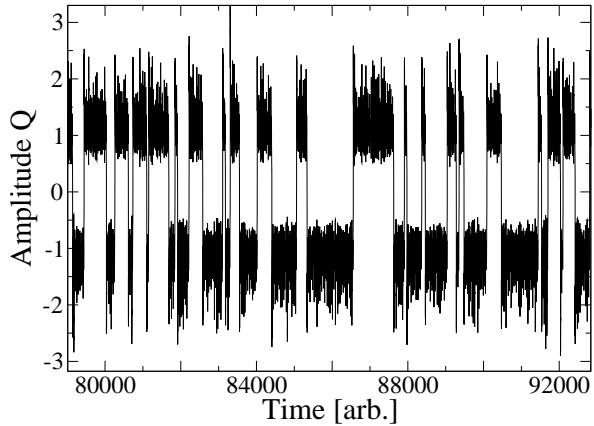}
\end{center}
\caption{Numerical integration of the amplitude equations (2). Time recording of the
amplitude of the quadrupolar mode for $\mu = 0.119$, $\nu=0.1$ and $\Gamma = 0.9$. (Figure from Gissinger \textit{et al.} 2010).}
\label{timeDQV}
\end{figure}

The dynamical system (2) with $\Gamma = 0$ occurs in different hydrodynamic problems, such as  the dynamics of periodic structures in thermoaline convection. It has been analysed in detail (Hughes \& Proctor 1990). The relative signs of the coefficients of the nonlinear terms have been chosen so that the solutions do not diverge when $\mu > 0$ and $\nu < 0$. Their modulus can be taken equal to one by appropriate scalings of the amplitudes. The velocity mode is linearly damped and its linear coefficient can be taken equal to $-1$ by an appropriate choice of the time scale. Note that similar equations were obtained through drastic truncation of MHD equations (Nozi\`eres 1978). However, in that context $\mu$ should be negative and the damping of the velocity mode is discarded, thus strongly modifying the dynamics.

This system displays reversals of the magnetic modes $D$ and $Q$ for a wide range of parameters. A time recording is shown in figure~\ref{timeDQV}. The mechanism for these reversals results from the interaction between the modes $D$ and $Q$ coupled by the broken ${\cal R}_{\pi}$ symmetry when $V \neq 0$. It is thus similar to the one described above but also involves an important difference: keeping the damped velocity mode into the system generates chaotic fluctuations. It is thus not necessary to add external noise to obtain random reversals. This system is fully deterministic as opposed to the one of P\'etr\'elis and Fauve (2008). Its phase space shows the existence of chaotic attractors in the vicinity of the $\pm {\bf B}$ quasi-stationary states. When these symmetric attractors are disjoint, the magnetic field fluctuates in the vicinity of one of the two states $\pm {\bf B}$ and the dynamo is statistically stationary.  When $\mu$ is varied, these two attractors can get connected through a crisis mechanism (Grebogi \textit{et al.} 1982), thus generating a regime with random reversals.  Compared to previous deterministic models, the one obtained here displays dynamical and statistical properties that are much closer to the ones of the direct simulations at low $P_m$ of Gissinger  \textit{et al.}  (2010) and  of the VKS experiment. For instance, the direct recordings of $D$ or $Q$ do not involve the growing oscillations characteristic of reversals displayed by the Rikitake (1958) or Lorenz (1963) systems but absent in dynamo experiments or in direct simulations. Correspondingly, the probability density function of $D$ is also much closer to the one obtained in experiments or direct simulations than the one of previous deterministic models.

\section{Large scale circulation over a two dimensional turbulent flow}\label{Kolmogorov}

\subsection{Large scale quasi-two-dimensional flows}\label{2D}

Rapid rotation or an externally applied strong magnetic field  have been known since a long time to make the flow two-dimensional by inhibiting velocity gradients along the direction of the rotation axis or of the magnetic field (Chandrasekhar 1961). This situation is  of  obvious interest for geophysical or astrophysical flows for which rotation can be important. The properties of turbulent flows strongly differ between these nearly two dimensional (2D) systems and the usual three dimensional ones.  
In 2D, it  is well known that an inverse cascade tends to accumulate energy in the largest scales of the system. 
Early studies of periodically driven flows (Sommeria 1986, Tabeling {\it et al.} 1987) have shown that an array of vortices becomes unstable when the forcing is increased and that a large scale  circulation is generated in the whole system. This large scale circulation is expected to display  rich dynamics because it is coupled to the smaller scale turbulent fluctuations.  Both the large scale field and the fluctuations are therefore components of the same turbulent  field, the velocity field.  This differs from the the former example in which reversals involve a field (the magnetic field) that is different from the  turbulent velocity field. 

\subsection{Reversals of the large scale circulation in Kolmogorov flows: experiments and numerical simulation}\label{kolmogorov reversals}

Following Sommeria (1986) and with the purpose of studying the  dynamics of the large scale circulation, we consider a square layer of mercury of length $L=12$ cm and depth $a=1.6$ cm. This layer is placed into a vertical magnetic field up to $B=500\,G$. An electric current $I$ up to $40$ A is injected in the fluid through $36$ electrodes placed on a periodic square array.  This current generates a Laplace force which drives the motion of the fluid. When  the applied magnetic field is large enough, the flow is mostly two dimensional. 
At low current, a flow made of $36$ counterrotating vortices is generated. This velocity field, obtained from direct  numerical simulations presented  below,  is shown in fig. \ref{figbas1} (a). At larger current, a large scale flow is generated.  Two states are possible and are shown in fig. \ref{figbas1} (b,c). Let  $(x,y)$ be the coordinates measured from the lower left corner of the square.   We note that the forcing is invariant under the reflections with respect to the planes  $x=L/2$ and $y=L/2$. We call  $\mathcal{S}_{x}$  (respectively $\mathcal{S}_{y}$) these symmetries. The large scale circulation thus  breaks the symmetry of the forcing by selecting a direction of rotation. 

\begin{figure}[!htb]
\begin{center}
\includegraphics[width=16cm,height=8cm]{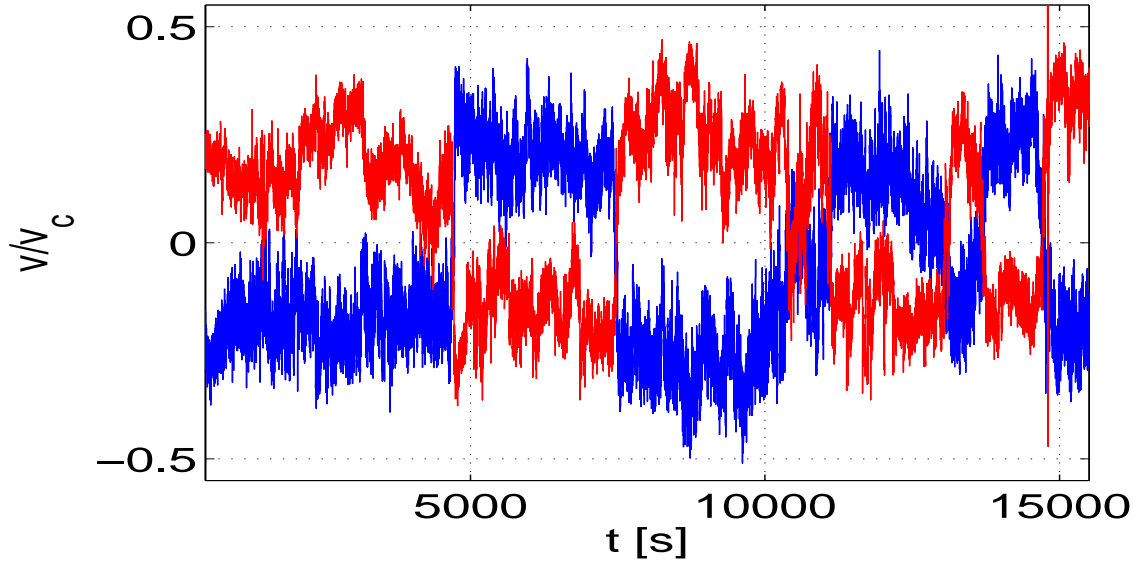}
\end{center}
\caption{Time series of the velocity $V/V_c$ where $V_c=(B I/(\rho a))^{1/2}\simeq4.28\,cm.s^{-1}$ for $Rh=37.7$ and $Re=45000$. The two curves correspond to measurements located on opposite sides of the cell. The velocity is averaged between  one centimeter and three centimeters away from the lateral wall.}
\label{johann}
\end{figure}

For large Reynolds number $Re=(B I L^2/(\rho a  \nu^2))^{1/2}$,  there exists a critical  value of $Rh=(a I/(B \sigma \nu L^2))^{1/2}$ under which the flow displays apparently random reversals between the two possible large scale circulations (Sommeria 1986).  We present measurements of the $x$ component of the velocity at position $(6; 2)$ cm and $(6; 10)$ cm. The velocity is measured using Vives probes and is averaged over $2$ cm.  In fig. \ref{johann}, a time series is displayed in the regime of random reversals.
The direction of the large scale circulation remains constant for long durations, here larger than one hour, and reverses in less than $100 s$. 
Note that the shape of the reversals does not display a clear overshoot.  In the case of the VKS experiment, reversals are observed when the two disks rotate at different speeds. Then the ${\cal R}_{\pi}$ symmetry is broken and this constrains the form of the phase space (fig. \ref{saddlenode}) which enforces the shape of the reversals and the existence of an overshoot. In contrast, in the case of the 2D turbulent flow,  no externally broken symmetry imposes an overshoot. 

A numerical simulation of this experiment has been performed. When the magnetic field is strong enough, the flow is almost 2D and the stream function $\psi$ obeys the two-dimensional Navier-Stokes equation
\begin{equation}
\partial_t \Delta \psi + J(\Delta \psi,\psi) = - \frac{1}{Rh} \Delta \psi + \frac{1}{Re} \Delta^2 \psi + 6 \pi \sin (6 \pi x) \sin (6 \pi y) \,. 
\label{NS}
\end{equation}
The first term on the right hand side is linear friction coming from the Hartmann layers on the bottom of the cell. The second one is the viscous term, and the last one mimics the electromagnetic forcing generated by the $6 \times 6$ electrodes.
We solved this equation using a pseudo-spectral code with uniform time-stepping and stress free boundary conditions (that is vanishing normal velocity and tangential constraint at the boundaries). The stream function is decomposed in the domain $(x,y) \in [0;1]^2$ on the basis $\left(\sin(n\pi x) \sin(p\pi y)\right)_{(n,p)\in\{1,...,N\}^2}$. The linear terms are computed in Fourier space, while the nonlinear one is computed in real space. All the numerical runs described further were performed using $256$ Fourier modes in each direction. We checked convergence: doubling this number of modes does not modify the numerical solution.


\begin{figure}
\begin{center}
a.\includegraphics[width=5cm,angle=0]{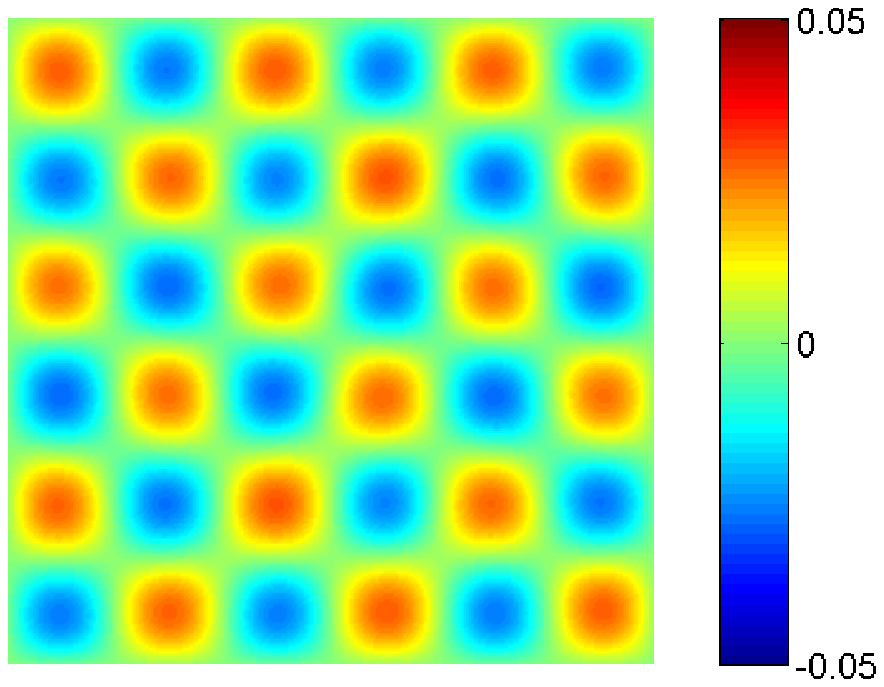} b.\includegraphics[width=5cm,angle=0]{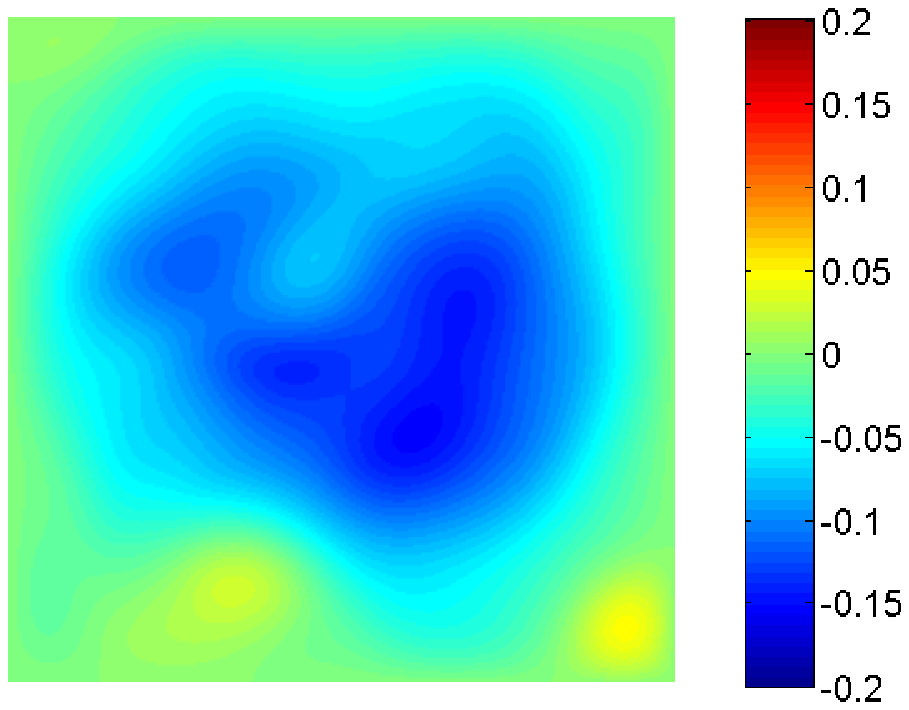} c.\includegraphics[width=5cm,angle=0]{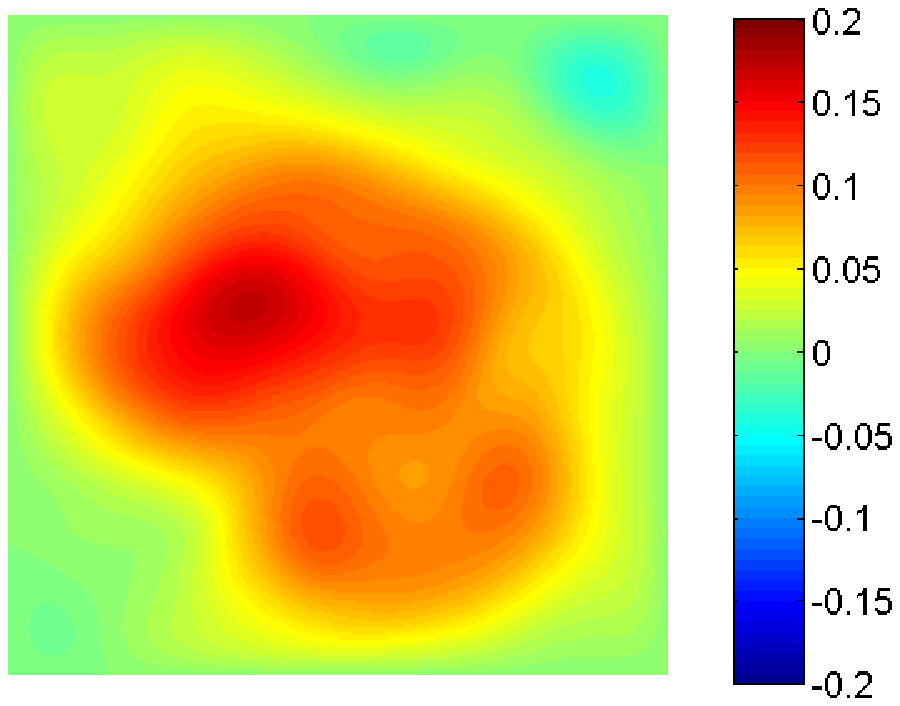}
\\
\caption[]{Stream function of the numerically computed solution of equation \ref{NS} for $Re=2000$, $Rh=55$,  a. Snapshot at short time, the velocity field displays the same periodicity as the forcing. b and c are two snapshots  displaying  large scale circulations of opposite signs. The large scale flow has broken symmetries of the forcing. Under certain conditions, the  circulation  reverses from one state to the other.}
\label{figbas1}       
 \end{center}
\end{figure}

\begin{figure}
\centering
\includegraphics[width=10cm,height=6cm,angle=0]{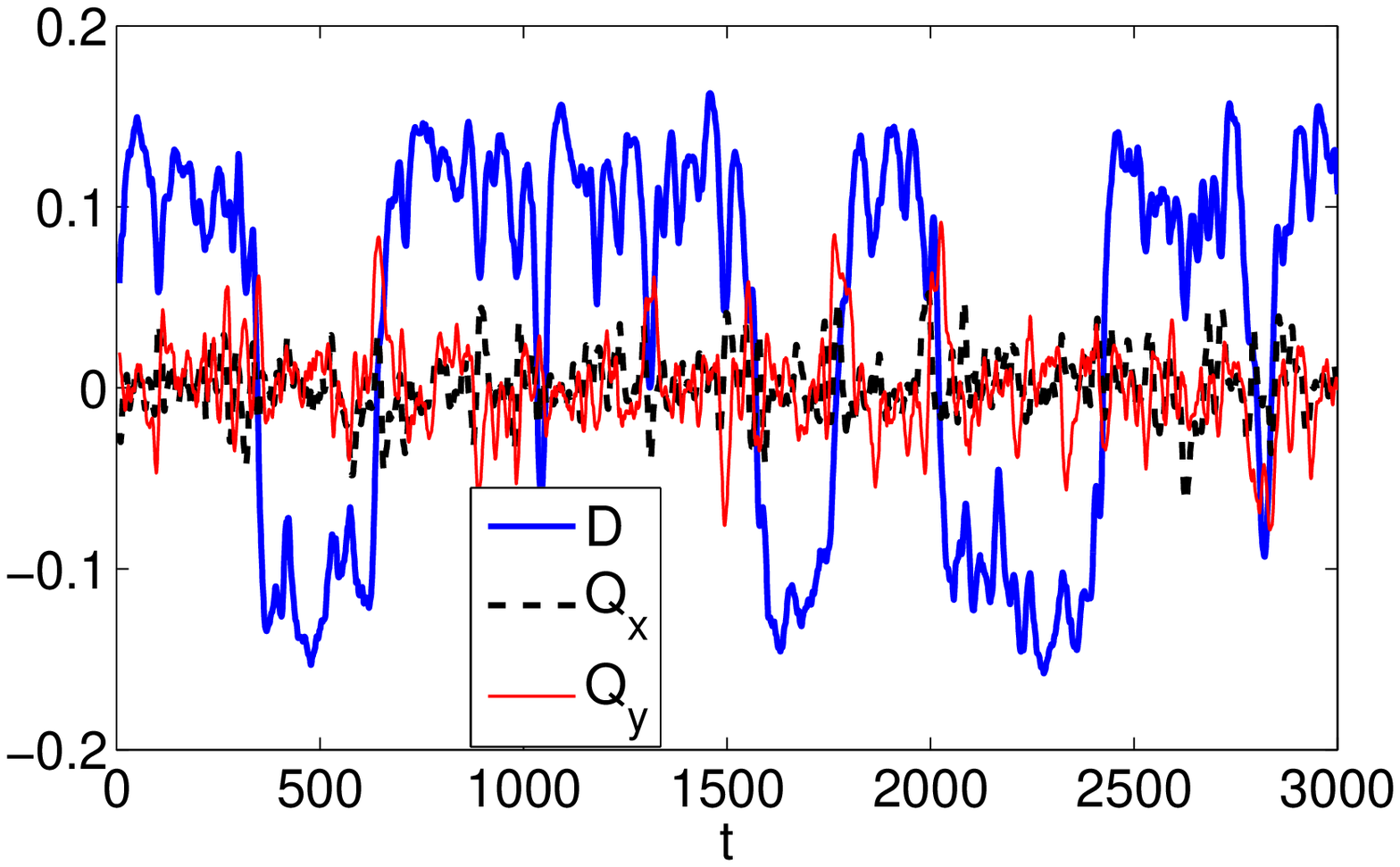}
\includegraphics[width=6cm,height=6cm,angle=0]{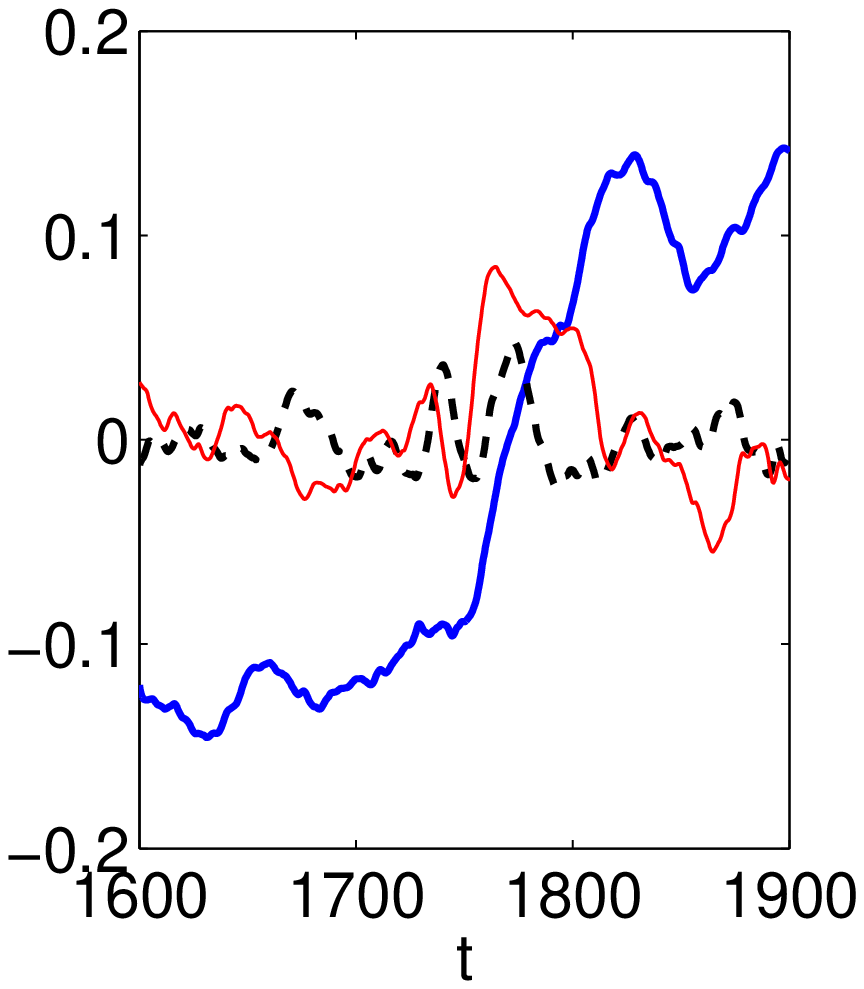}
\\
\caption[]{Left: Time series of the amplitude of the large scale circulation $D$ and of two large scale components $Q_x$ and $Q_y$ (see definition in the text) obtained numerically for $Rh=55$ and $Re=2000$. Right: zoom on one reversal. A sliding average is applied to reduce high frequency fluctuations.}
\label{figbast}       
\end{figure}

%

\subsection{A dynamical system displaying reversals of the large scale circulation}\label{DQxQy}

Reversals of the large scale circulation are  observed in this direct numerical simulation. An example of time series is shown in fig. \ref{figbast}.  It is possible to identify the large scale structures that are involved in the process. They can be decomposed into three main components :
\begin{itemize}
\item The amplitude of the large-scale circulation is denoted as $D$ and is defined here as  the Fourier coefficient of  $\sin(\pi x) \sin(\pi y)$ in the decomposition of the streamfunction.
\item A second component, called $Q_x$, is  the part of the large scale flow that is quadrupolar in the $x$ direction. It is even under $\mathcal{S}_{x}$  and odd under $\mathcal{S}_{y}$. It can be measured by the  Fourier coefficient of $\psi$ in $\sin(2 \pi x) \sin(\pi y)$.
\item A third component, called $Q_y$, is  the part of the large scale flow that is quadrupolar in the $y$ direction. It is odd  under $\mathcal{S}_{x}$  and even  under $\mathcal{S}_{y}$. It can be measured by the  Fourier coefficient of $\psi$ in $\sin( \pi x) \sin(2 \pi y)$.
\end{itemize} 
We  observe that reversals of $D$ take place when $Q_y$ reaches large values.   
In agreement with the experiment, no overshoot is seen at the end of the reversals. 

In order to obtain the minimal model that displays  similar   dynamics, 
we note that the three components change differently under the different symmetries of the system. As already mentionned $D$ and $Q_x$ are odd under $\mathcal{S}_{y}$  while $Q_y$ is even.  $D$ and $Q_y$ are odd under $\mathcal{S}_{x}$ while $Q_x$ is even.  A simple dynamical system that verifies  these symmetries is 
\begin{eqnarray}
\dot{D} & = & -\nu D-Q_x Q_y \nonumber \\
\dot{Q_x} & = &  Q_x-Q_y D-Q_x^3  \nonumber \\
\dot{Q_y} & = & \mu Q_y+D Q_x \,.
\label{proc}
\end{eqnarray}

\begin{figure}[!htb]
\begin{center}
\includegraphics[width=14cm,height=8cm]{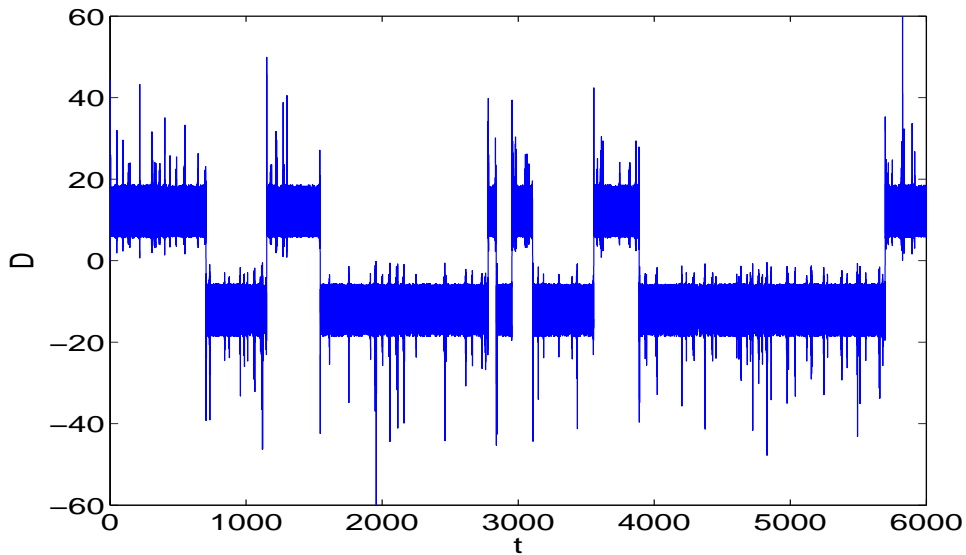}
\end{center}
\caption{Time series of $D$ solution of equations (\ref{proc}) in a regime of reversals ($\mu=6$ and $\nu=2.65$).}
\label{modelebasile}
\end{figure}

Compared to Hughes and Proctor (1990), the set of equations (\ref{proc}) contains an extra third-order nonlinearity in the last equation. A time series is represented in figure \ref{modelebasile} for $\mu=6$ and $\nu=2.65$: the large scale circulation $D$ reverses randomly. As in the direct  numerical simulations, the value of $Q_y$ increases during the reversals of $D$. 

One should notice that the present problem maps onto the magnetic one in the following way: if we replace $D$, $Q_y$, and $Q_x$ respectively by the amplitude of the dipolar magnetic component $D$, quadrupolar magnetic component $Q$, and equatorially-antisymmetric velocity component $V$, the symmetries  $\mathcal{S}_{x}$ and $\mathcal{S}_{y}$ lead to the same constraints on the dynamical system than the invariance ${\bf  B} \rightarrow {\bf - B}$ and the equatorial symmetry for the magnetic field of an astrophysical object. Therefore the system of equations (\ref{proc}) could describe the magnetic field of an astrophysical object as well, and leads to reversals of the magnetic dipole with no imposed equatorial-symmetry breaking. In this sense it is  slightly different from model (2)  in which the dynamics of $V$ (corresponding to $Q_x$) results from an externally imposed broken symmetry and not from an instability of the large scale velocity field that spontaneously breaks this symmetry. 
In these low dimensional models, the system follows chaotic trajectories that wander in the basin of attraction associated to one polarity of the large scale field or its opposite. In the regime of reversals, the chaotic attractor connects the two basins of attraction and after a long duration the system escapes from one state toward its opposite.

\section{Large scale circulation in turbulent Rayleigh-B\'enard convection}\label{convection}

\subsection{Experiments on the dynamics of the large scale circulation in Rayleigh-B\'enard convection}\label{LSC}

Rayleigh-B\'enard convection is a canonical example of cellular instability. It is achieved by uniformly heating
from below a horizontal layer of fluid. For small temperature  difference $\Delta T$ across the layer of depth $d$, the fluid remains in a stable heat conducting state, with a linear temperature profile and no fluid motion. However, if the fluid's density is a decreasing function of temperature,  the thermal gradient generates a density stratification with cold heavy fluid above warm light fluid. For sufficiently large temperature differences, the resulting buoyancy force overcomes dissipative effects due to kinematic viscosity, $\nu$, and heat diffusivity, $\kappa$, causing less dense warmer fluid to rise and cooler fluid to sink. When the top and bottom boundaries have a thermal conductivity much higher than the one of the fluid, i.e. their temperatures are homogenous, periodic parallel convection rolls of horizontal size comparable to $d$, result from the circulation of the fluid, when the Rayleigh number, $Ra=\alpha g \Delta T d^3/(\nu \kappa)$, is larger than a critical value, $Ra_c$ ($g$ is the acceleration of gravity, $\alpha$ is the coefficient of thermal expansion of the fluid). In the case of a large aspect ratio container (ratio of the horizontal size $L$ to the depth $d$), it has been observed by Krishnamurti and Howard (1981) that a large scale horizontal shear flow of horizontal size $L$ is generated for large enough values of $Ra$, roughly three orders of magnitude larger than $Ra_c$. In contrast to the large scale flow generated by the Kolmogorov forcing considered in section~\ref{Kolmogorov}, the dynamics of this flow has not been studied. For convenience, most high Rayleigh number experiments have been performed in cylindrical containers of aspect ratio one. In that case, the large scale circulation evolves out of the cellular structure observed at convection onset without implying a secondary bifurcation and an associated broken symmetry. However, even in that simple case, this large scale circulation displays dynamics that have been first related to flow reversals (see Ahlers \textit{et al.} 2009 for a review of the early experiments and related models). It has been realised that, although simpler for the machine shop, the cylindrical geometry is not appropriate to study flow reversals. The convective pattern at onset and thus the large scale circulation for large $Ra$ indeed break the rotational invariance of the container.  Thus, the dynamics primarily results from the azimuthal orientation of the large scale circulation and the flow measured locally can vanish without necessarily implying a cessation of the whole large scale circulation (Ahlers \textit{et al.} 2009). Both reorientations and cessations of the large scale circulation have been evidenced by recent direct numerical simulations (Mishra \textit{et al.} 2010).  They also displayed a quadrupolar temperature distribution during flow cessations (as opposed to the dipolar distribution resulting from the large scale circulation). Thus, although different mechanisms are superimposed in the dynamics of large scale circulation in cylindrical containers, flow cessations involve the competition between two large scale modes with different symmetries, in striking analogy with the mechanism of reversals of the magnetic field proposed in \ref{model}.

Only a few studies of the dynamics of the large scale circulation have been performed in parallelepipedic containers. Liu and Zhang (2008) reported a large scale flow in an elongated box of aspect ratio $2.6$ for $Ra > 10^7$ together with reversals but they provided a detailed study only with freely moving nylon spheres within the flow. Another experiment has been performed recently in a quasi-two-dimensional parallelepipedic container of aspect ratio one, and its results have been compared to direct numerical simulations of two-dimensional convection (Sugiyama \textit{et al.} 2010). Although the authors insist on the importance of corner flows superimposed to the large scale diagonal circulation, a more pertinent observation is related to the change of symmetry of the large scale flow observed during the reversal process. When the dipolar part of the large scale flow vanishes, a quadrupolar structure is clearly visible (see their figure 1b). We also note that earlier two-dimensional numerical simulations (Breuer \& Hansen 2009) performed in a container of aspect ratio two, also displayed the competition between two large scale modes with different symmetries during the reversal process.  Thus, it appears that, as for reversals of the magnetic field in the VKS experiment, reversals of the large scale convective flow in the quasi-two-dimensional configuration also involve the competition between two large scale modes with different symmetries. Another similarity between the convective and dynamo reversals is related to their occurrence in a finite interval range of the control parameters  (Rayleigh and Prandtl numbers $Ra$, $Pr$ or magnetic Reynolds numbers  $R_m$). In both two-dimensional numerical simulations of convection, it has been also reported that relatively small changes in the aspect ratio strongly modify the reversal dynamics and thresholds. This probably results from the inhibition of the mode with quadrupolar symmetry that is involved in the reversal process. Finally, the numerical simulations of Sugiyama \textit{et al.} (2010) show that reversals disappear when the Prandtl number of the fluid is  too small ($Pr \leq 0.7$) or too large. We will report below reversals in a cubic container of mercury ($Pr \simeq 0.025$).
Applying a horizontal magnetic field enables to make the flow more and more two-dimensional by inhibiting velocity gradients along the direction of the magnetic field (Fauve \textit{et al.} 1984). We will show that this strongly affects the modes involved in the reversal process.

\subsection{Dynamics of the large scale circulation in a cubic container subject to a horizontal magnetic field}

\begin{figure}[!htb]
\begin{center}
a.\includegraphics[width=8cm]{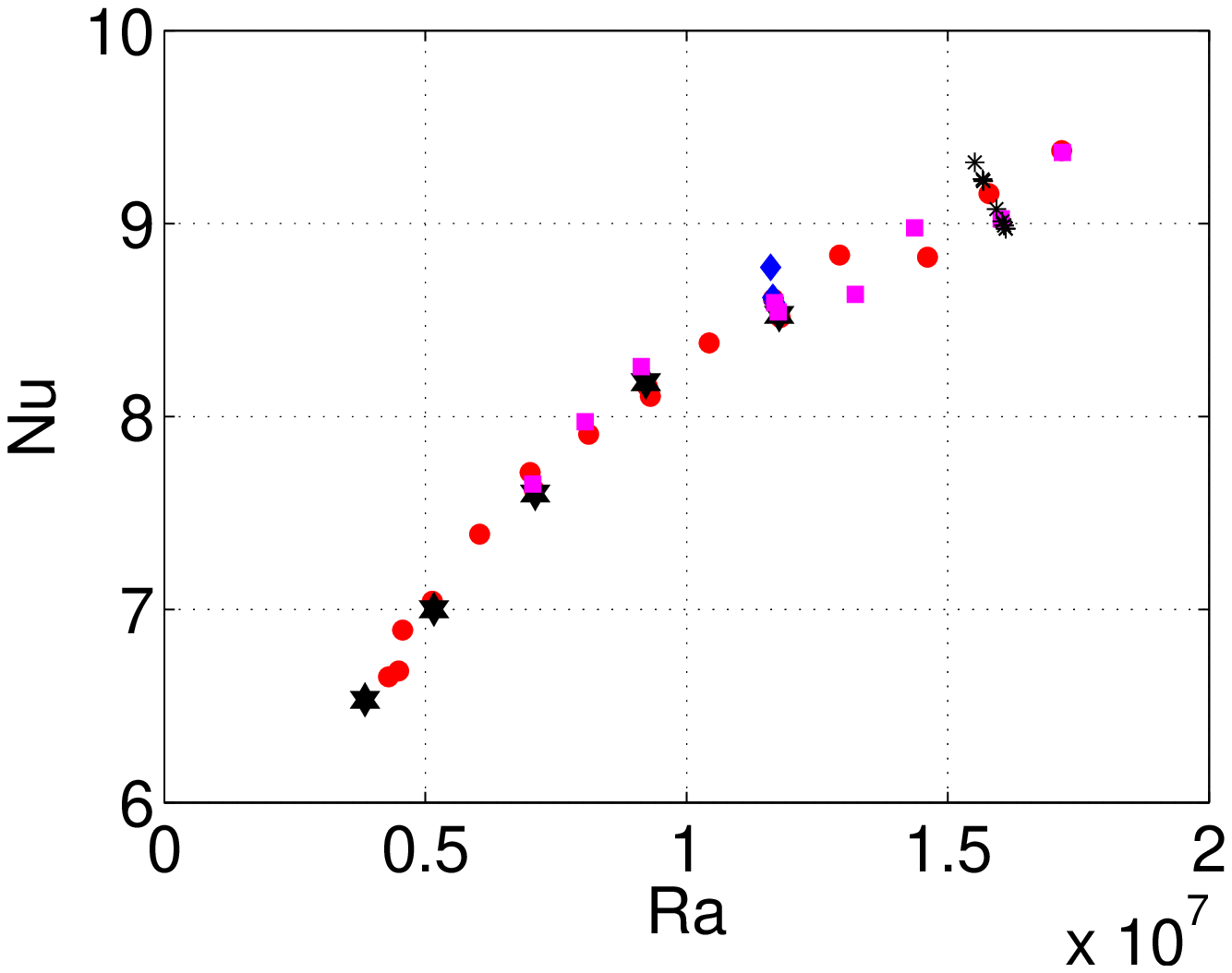}b.\includegraphics[width=8cm]{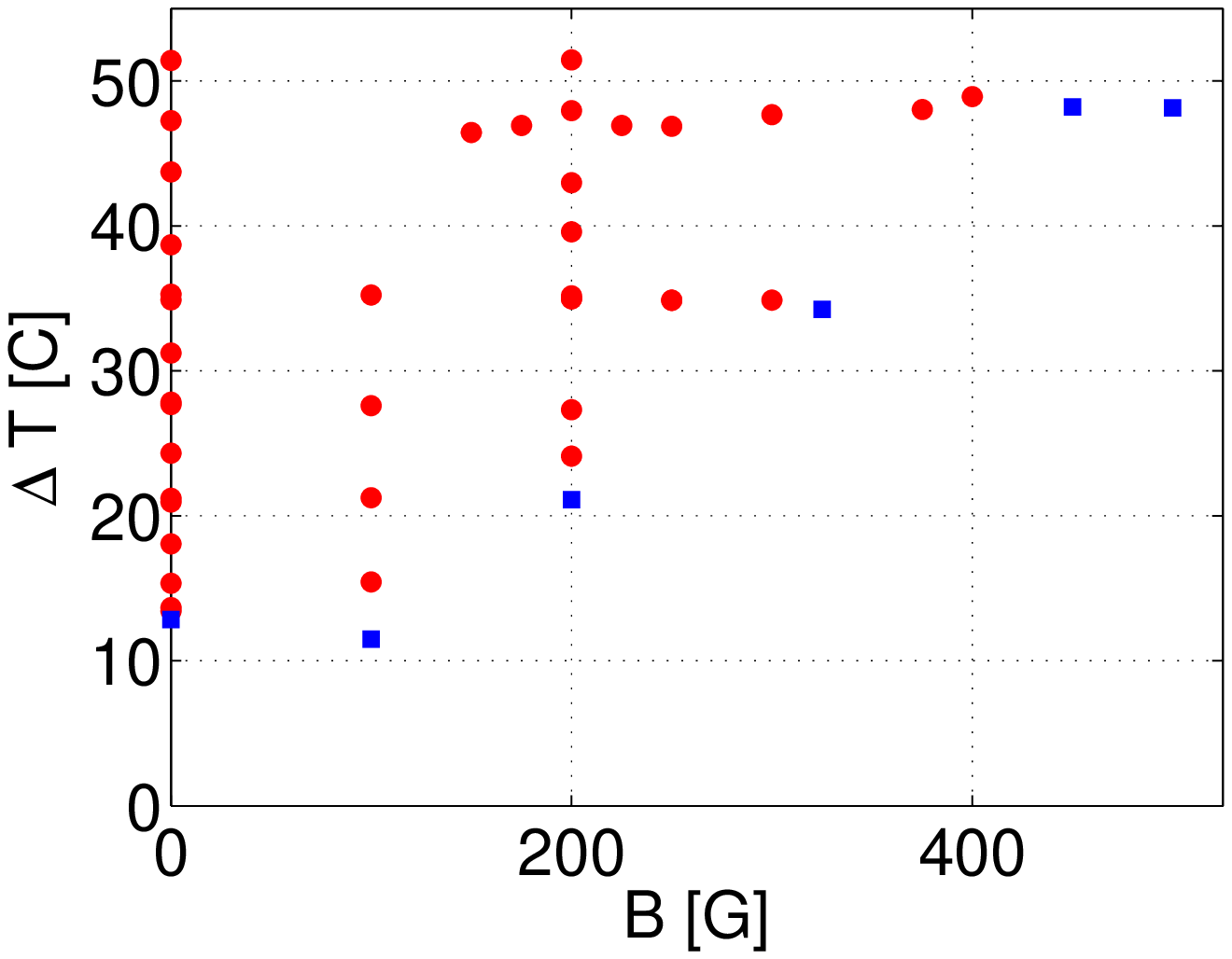}
\end{center}
\caption{a. Nusselt number as a function of the Rayleigh  number for ($\circ$): $B=0\, G$,   ($\star$): $B=100\, G$, ($\square$): $B=200\, G$, ($*$): $B$ varies from $0$ to $500 \,G$ at $\Delta T\simeq 48^o C$, ($\diamond$): $B$ varies from $0$ to $325 \,G$ at $\Delta T\simeq 35^o C$. b. Parameter space ($B, \Delta T)$. The blue squares correspond to stationary regimes. The red circles correspond to reversals of the large scale circulation.  
\label{NuvsRa}}
\end{figure}

We consider a cubic container  of size $d=45$ mm filled with mercury of density $\rho=13.5\, 10^3\, kg.m^{-3}$,  kinetic viscosity $\nu=1.12\, 10^{-7} m^2.s^{-1}$, thermal diffusivity $\kappa=4.3\,10^{-6} m^2.s^{-1}$, specific heat $c_p=138 J.kg^{-1}.K^{-1}$ and coefficient of thermal expansion  $\alpha=1.8\,10^{-4} K^{-1}$. The side walls of the experiment are made of PVC while the  top and bottom ones are copper plates. Water circulation through the upper copper plate maintains the upper  temperature fixed. A constant heat flux $F$ is imposed at the bottom boundary. The lower plate is covered with temperature probes which  allows to determine the temperature difference between the bottom and the top plate $\Delta T$ and also to identify the spatial structure of the temperature field in the experiment. 
A horizontal magnetic field, $B$, up to $500\,G$ can be applied perpendicular to one of the sides of the cell.  In fig. \ref{NuvsRa}, we display the Nusselt number $Nu=F/(\rho c_p  \kappa \Delta T d)$ as a function of the Rayleigh number. 
In this regime of parameters, the magnetic field has little effect on the heat transport. For $\Delta T\simeq 48^o C$ {\it i.e.} $Ra=1.58\,10^7$, the variation of the Nusselt number is less than $4 \%$ when $B$ is increased from $0$ to $500 \,G$ corresponding to an interaction parameter, $N \sim \frac{\sigma B^2}{\rho}\sqrt{\frac{d}{\alpha g \Delta T}}$, smaller than $0.2$. \\

In the absence of magnetic field, for $Ra$ smaller than $Ra_{r} \simeq4.3\,10^6$, the temperatures fluctuate around some constant value.  In this regime, the larger scale of the convective pattern is a single roll that fills the whole cell. This roll is expected to be perpendicular to one of the sides of the cell and we have checked that its direction  can be rotated by tilting the cell. However, no spontaneous change in the large scales of the temperature pattern is observed even for  measurement durations larger than  $24$ hours.

\begin{figure}[!htb]
\begin{center}
\includegraphics[width=18cm]{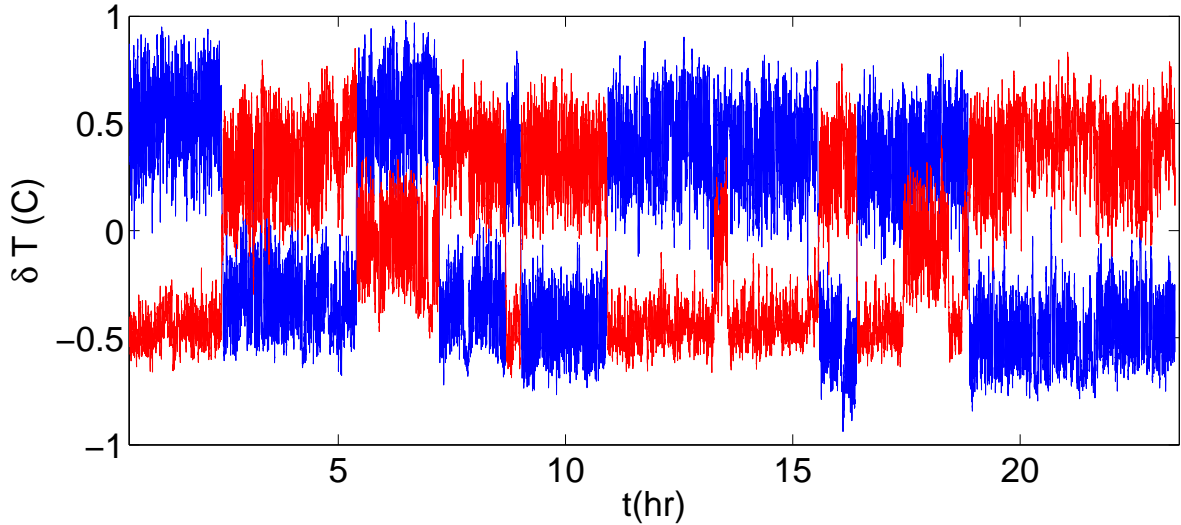}
\end{center}
\caption{Time series of the temperature measured at two opposite corners of the lower plate for $B=0$, $\Delta T=15.4^o C$ and $Ra=5.13\,10^6$. Measurement rate is $1/3$ Hz and a sliding average over $10$ measurements is performed to reduce fluctuations.}
\label{Tvst}
\end{figure}

A change of behaviour is observed when $Ra$ reaches $Ra_{r}$. The temperatures fluctuate for long duration around some value but from time to time, the whole temperature pattern  changes suddenly. Two time series of the temperature measured at opposite positions on the diagonal of the cell, $40$ mm away from the corners, are displayed in fig. \ref{Tvst}. These large scale modifications  are separated by very long time intervals, so that statistical properties can only be estimated. We have measured that the mean duration between these events decreases  from more than $3$ hrs to around  $30$ minutes when $\Delta T$ changes from $13.4$ to $27.7^o\,C$.

\begin{figure}[!htb]
\begin{center}
a.  
\includegraphics[width=8cm]{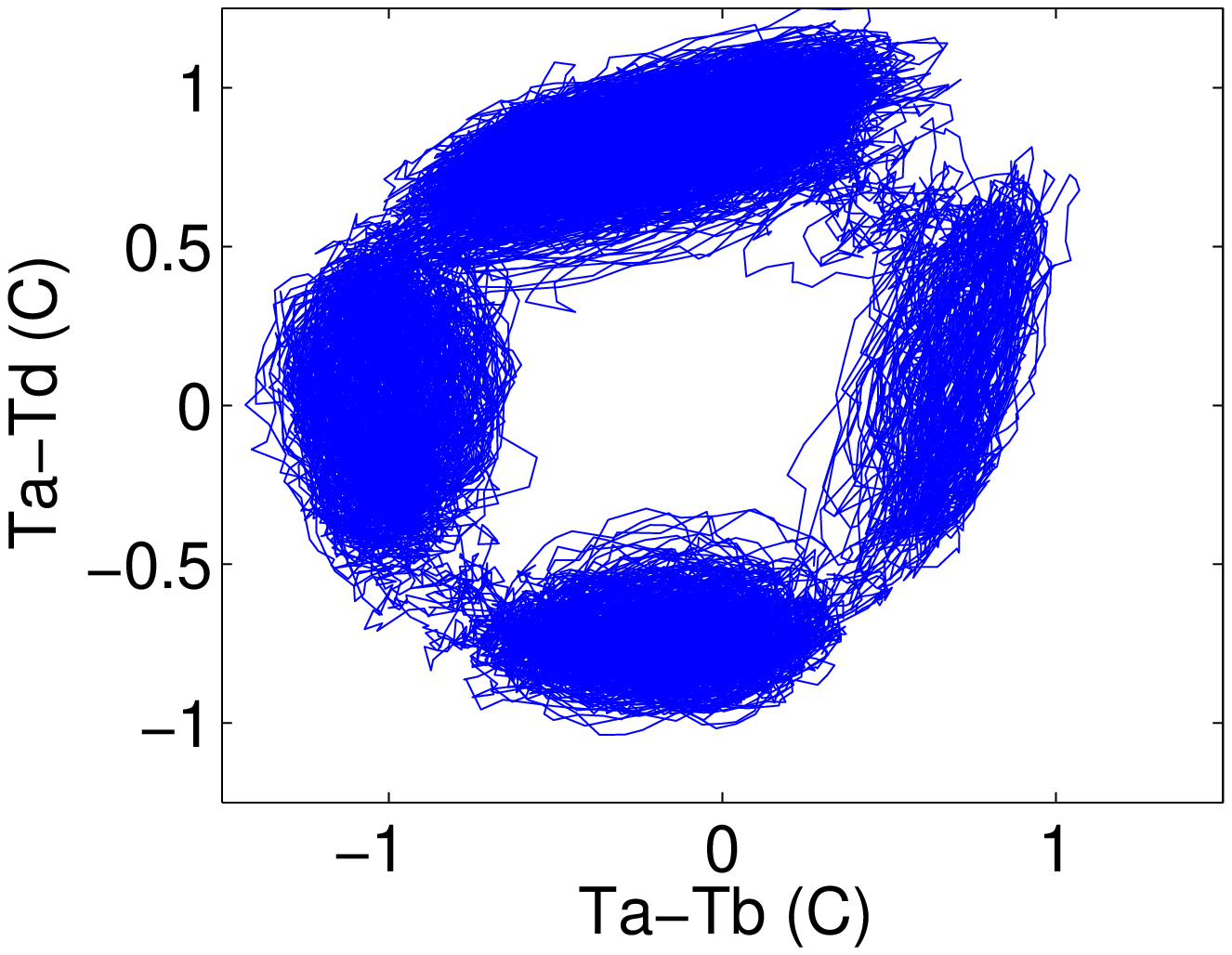}\\
b.  
\includegraphics[width=8cm]{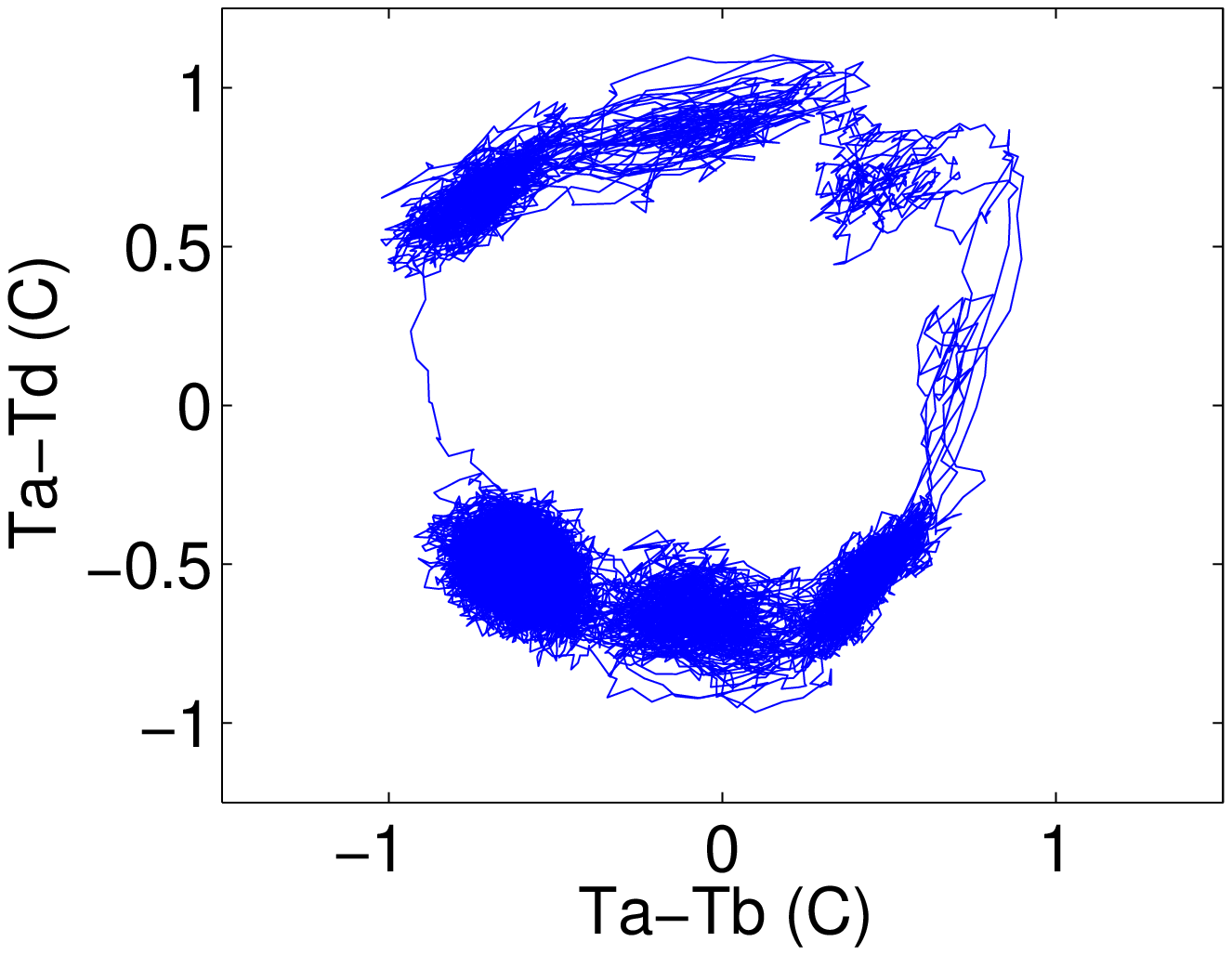}
\end{center}
\caption{Projection of the phase space obtained from temperature differences: $T_a-T_d$ and $T_a-T_b$. a- In the absence of applied magnetic field, same data as fig. \ref{Tvst}. b- For  $\Delta T= 48^o C$,  a  magnetic field $B=375\,G$ is applied parallel to the direction $a-b$.}
\label{figphase}
\end{figure}

Although the time series share common features with the ones of the reversals observed in the VKS experiment  (see fig. \ref{figvks}) or in the two dimensional turbulent Kolmogorov flow (see fig. \ref{johann}), we note that the evolution is more complicated: there are more than two stationary states and we can identify events where one of the temperatures changes with no noticeable evolution of the other one. Careful inspection of the evolution of several probes shows that the modifications of the large scale of the flow  are associated to a rotation of the large scale roll.  This behaviour can be identified by considering the horizontal temperature gradients. In fig. \ref{figphase},  we plot a cut in  phase space $(T_a-T_b,\,T_a-T_d)$ where $T_a-T_b$ and $T_a-T_d$ are temperature differences measured along two perpendicular sides of the cell.   The trajectories in phase space are roughly located along a  circle. Between the changes of direction, the system fluctuates close to one of the four states that correspond to the roll rotation vector pointing towards one of the four lateral sides of the cell. Indeed if a roll is aligned with the line $a-b$, then $T_a-T_b$ is roughly zero and the direction of rotation of the roll is given by the sign of $T_a-T_d$. 
The trajectory in phase space shows that the roll  evolves from being perpendicular to one of the sides to being perpendicular to one of the two neighbouring sides. Its direction thus changes by  $\pm 90^o$. In other words, the large scale flow displays random changes of direction and reversals are obtained by the sequel of two changes of orientation in the same direction, which results in a rotation of $180^o$.


\begin{figure}[!htb]
\begin{center}
a. 
\includegraphics[width=8cm]{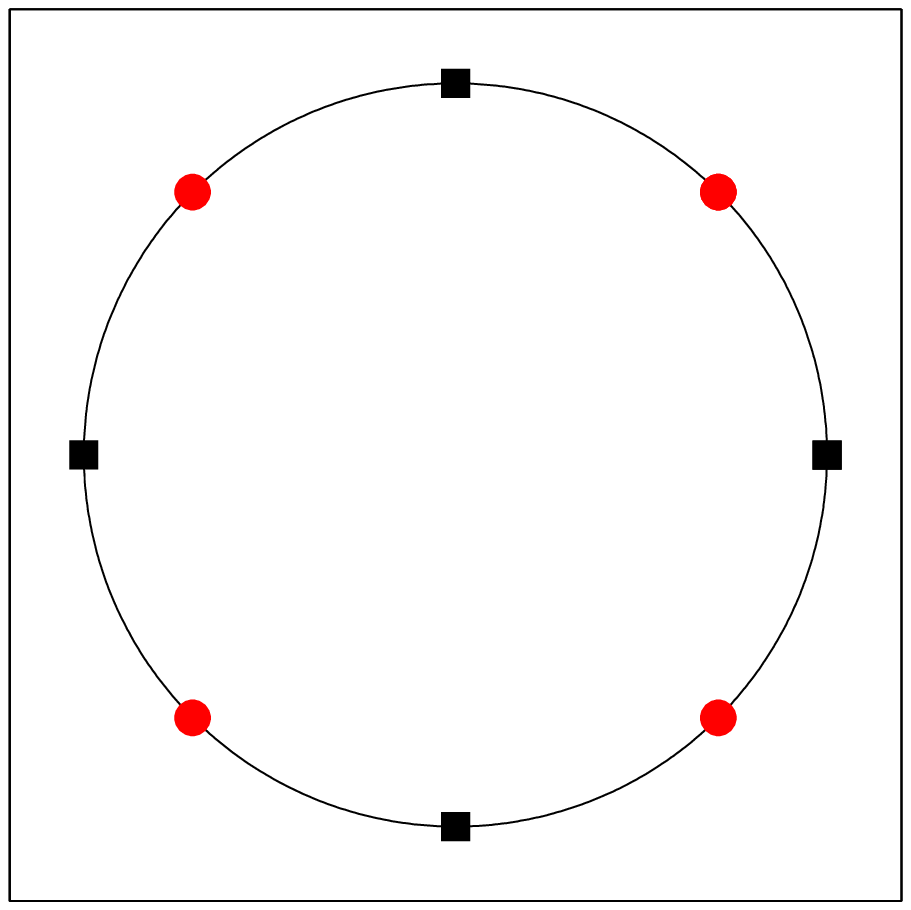}\\
b. 
\includegraphics[width=8cm]{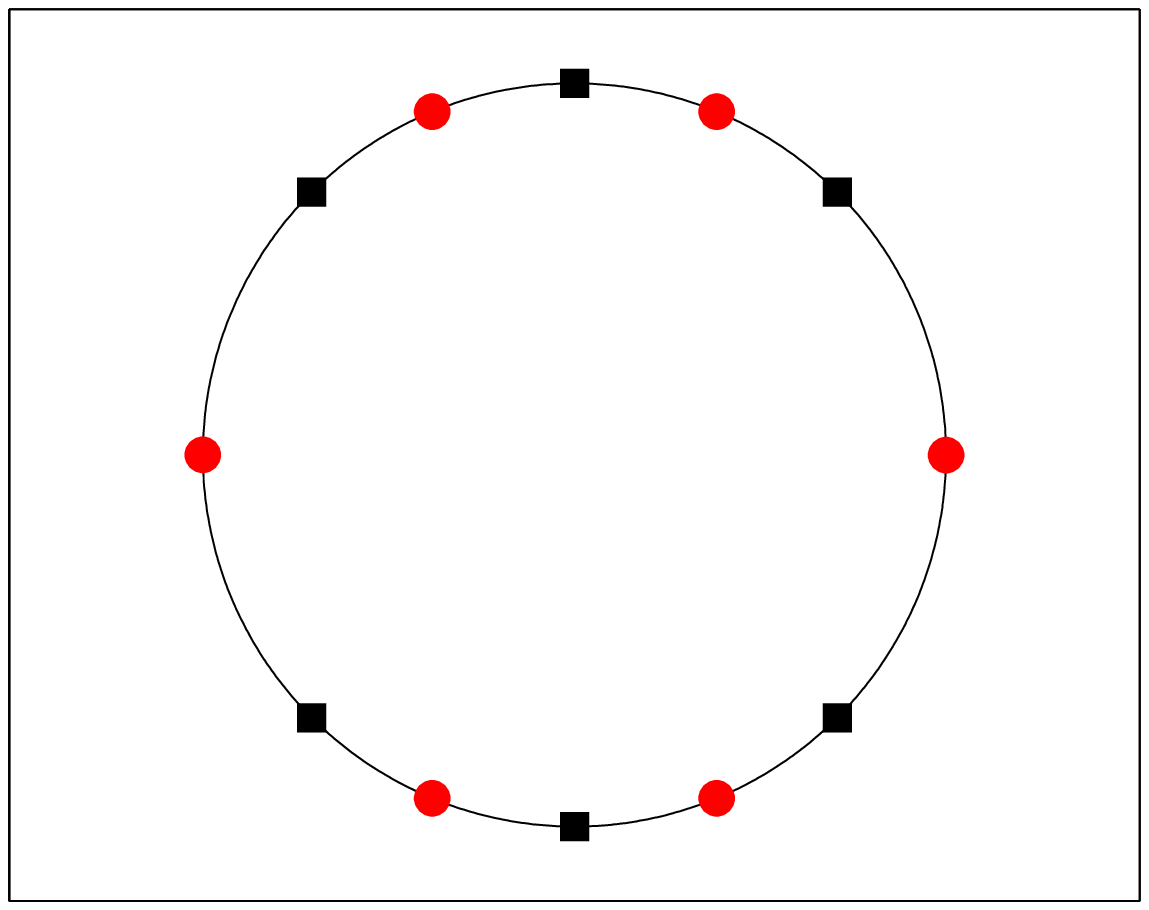}\\

\caption{Phase space displaying stable solutions ($\square$ black) and unstable solutions ($\circ$ red). a- The four-fold symmetry is the one of the experiment in the absence of magnetic field. b-  In the case of an applied magnetic field, the two formerly stable solutions representing rolls  perpendicular to the field are unstable and two pairs of new stable  solutions (the  diagonal rolls) are formed through a pitchfork bifurcation. Considering that turbulent fluctuations allow to evolve from a stable solution toward its neighbouring ones,  both phase spaces represented in fig. \ref{figphase} can be reconstructed.
\label{figmodelconv}}
\end{center}
\end{figure}

The obtained phase space clearly displays the four-fold symmetry of the experiment. A way to lower the degree of symmetry is to apply a horizontal magnetic field. Measurements have been performed in which a field is applied parallel to the line joining $a$ and $b$. In the parameter space displayed in fig. \ref{NuvsRa} b, we observe that reversals disappear when the magnetic field is large enough.  When reversals are observed in the presence of an applied magnetic field, 
the trajectories in phase space are  modified. As displayed in fig. \ref{figphase} b, the two states corresponding to rolls perpendicular to the field ($T_a-T_d\simeq 0$) have lost stability: the system spends only a short time in these states. In contrast the system spends long durations in the vicinity of four new states. These states correspond to $T_a-T_b\simeq \pm (T_a-T_d)$ which in physical space means that  a roll is aligned along the diagonal of the square. 
These observations result from  the applied  magnetic field  that reduces  the variations of the velocity  along the direction of the   field. The large scale dynamics is thus  modified by the following two effects: first, the rolls are preferably aligned along the magnetic field and rolls perpendicular to the field are expected to loose their stability; second, turbulence is reduced because  fluctuations along the direction of the field are damped. 
These two effects explain that the pair of  solutions perpendicular to the field turns unstable  and ultimately that large scale dynamics is suppressed.

From the point of view of low dimensional modelling, the regime with $B=0$ corresponds to a  system with four stable fixed points. In a planar model describing the competition between  perpendicular rolls, these stable points are located between four unstable states which are saddles (see fig. \ref{figmodelconv} a).  When a magnetic field is applied, a pair of stable solutions (transverse rolls) looses its stability. The four new states (diagonal rolls) that are observed can be described as two pairs of stable states that have been generated from a pitchfork bifurcation of the transverse rolls. The resulting phase space contains six nodes and, in-between, six saddles (see fig.  \ref{figmodelconv} b). A way to model the  random dynamics of the large scale flow is to consider that turbulent fluctuations act as noise that triggers transitions from one state to one of its neighbours. 

As in the case of reversals of the magnetic field, the large scale flow does not change sign directly under the influence of turbulent fluctuations (plumes, etc). This transition always involves another (or several other) large scale modes. In the case of square symmetry, perpendicular rolls are involved. When the square symmetry is broken by an external magnetic field, more complex mixed states are involved in the reversal process. For the Prandtl number of mercury, reversals are not observed above a critical value of the magnetic field. In contrast to the reversals of the magnetic field in the VKS experiment, an externally imposed broken symmetry is not needed to generate reversals of the large scale circulation in convection as well as in Kolmogorov flows.  Correspondingly, the reversals do not display an overshoot. 

\section{Discussion and conclusion}

We have presented three examples of turbulent flows in which a field generated at large scale reverses. Some properties are common to the three systems.
\newline  
In all cases the large scale field breaks a symmetry of the forcing and reversals allow the system to statistically recover this symmetry.  
\newline
A clear separation of time scales exists between the duration of a reversal and the waiting time between reversals. Indeed, random reversals are rare events that are initiated when the system wanders sufficiently far away from the basin of attraction of a fixed point. Once this phase has been completed, the evolution is fast because it does not require to reach a rarely visited domain in phase space.
 \newline
During a reversal the whole field does not vanish. In particular, it is possible to identify large scale structures that play a role in the dynamics and are visited by the system.

The three systems differ from the point of view of symmetries and in the way they are broken. The magnetic field and the large scale circulation have two stable states while the large scale roll in our convection experiment has four or six stable states. In the case of the experimental dynamo, one symmetry of the forcing is externally broken by rotating the two disks at different speeds. In the other two systems, one symmetry is spontaneously broken and this selects the direction of rotation  of the roll's axis or the sign of the transitional two cell flow. These symmetry properties are responsible for the differences in shape of the reversals observed in the three systems.   

The geometry of the large scale dynamics being governed by a low-dimensional dynamical system, we expect that reversals can also occur in laminar flows. This is indeed the case. Direct simulations of axisymmetric Rayleigh-B\'enard convection in a cylindrical geometry (Tuckerman \& Barkley 1988) as well as the weakly nonlinear analysis of the same system (Siggers 2003), have shown that the steady state observed above the convection onset can bifurcate to a travelling wave in which the number of rolls oscillates between two adjacent values. It has been emphasized that this transition to periodic behaviour is a symmetry-restoring bifurcation, as the states in the second half of the limit cycle are related to those of the first half by changing the sign of the velocity field (and accordingly of the temperature perturbation). A travelling spatially periodic pattern is indeed the simplest way to generate reversals of the flow measured at some location. This has been observed in direct numerical simulations of Rayleigh-B\'enard convection with horizontal periodic boundary conditions (Paul \textit{et al.} 2010). In that case, randomly occurring lateral shifts of the roll pattern by half a wavelength, lead to global flow reversals of the convective motion. These events are also observed in a finite interval range of Rayleigh number (although Ra is much smaller than in the presence of lateral boundaries). At first sight, one may consider that the presence of lateral boundaries also deeply modifies the qualitative bifurcation mechanism leading to flow reversals but it is worth considering this analogy further. 
Drift instabilities of cellular patterns have been indeed observed in various experiments, such as Couette flow between two horizontal coaxial cylinders with a partly filled gap (Mutabazi \textit{et al.} 1988),  film draining (Rabaud \textit{et al.} 1990),  directional crystal growth (Simon \textit{et al.} 1988), Faraday waves (Douady \textit{et al.} 1989), etc. They have been understood as a secondary instability that spontaneously breaks the reflection symmetry of the pattern which then travels in the direction related to this asymmetry (Coullet  \textit{et al.} 1989, Fauve  \textit{et al.} 1991, Caroli  \textit{et al.} 1992). If the reflection symmetry, say $x \rightarrow -x$, is externally broken, then one expects that a spatially periodic pattern that breaks the translational invariance along the $x$-axis, generically drifts along the $x$-axis at onset when the domain is infinite. This is easily understood in the framework of amplitude equations that amounts to describe a periodic pattern of wave number $k$ in the form $A(t) \exp ikx + c.c. + \cdots$. $A(t)$ is a complex amplitude, $c.c$ stands for the complex conjugate of the previous term and the dots represent higher order terms in the vicinity of the bifurcation threshold. It is well known that the form of the equation for $\dot A$ in series of successive powers of $A$ and $\bar{A}$ (the complex conjugate of $A$), results from the broken symmetries at the instability onset. If the system is invariant under the transformation $A \rightarrow - A$, this equation, up to leading order nonlinear terms, is similar to (\ref{eqdipquad})

\begin{equation}
\dot{A}=\mu A+\nu \bar{A}+\beta_1 A^3+\beta_2 A^2\bar{A}+\beta_3 A \bar{A}^2+\beta_4\bar{A}^3\,, \nonumber
\end{equation}
where all the coefficients are complex numbers. If the system is translationally invariant along the $x$-axis, the amplitude equation should be invariant by rotation in the complex plane, $A \rightarrow A \exp i \phi$, where $\phi$ is an arbitrary phase; thus $\nu = \beta_1= \beta_3 = \beta_4 = 0$. If the system has parity invariance, $x \rightarrow - x$, which implies the invariance $A \rightarrow \bar{A}$ for the amplitude equation, then all the coefficients should be real. Thus, we get the usual amplitude equation that describes the stationary bifurcation of a homogeneous periodic pattern,

\begin{equation}
\dot{A}=\mu_r A+\beta_{2r} \vert A \vert ^2\, A,
\end{equation} 
where all the coefficients are real.
When the parity invariance is externally broken, these coefficients generically have  non zero imaginary parts. When the imaginary part of $\mu$, say $\mu_i$, is non zero, then the pattern is generated through a Hopf bifurcation and  is thus a travelling wave that leads to periodic reversals of the flow at a given location.  
When the lateral boundaries are taken into account, translational invariance is externally broken and the leading order additional term that should be considered is $\nu \bar{A}$. This new term indeed tends to quench the phase of the pattern. The linear stability analysis of the solution $A=0$ gives the dispersion relation for the growth rate $s$ 
\begin{equation}
s^2 -2 \mu_r\,  s + \vert \mu \vert^2 -  \vert \nu \vert^2 = 0,
\end{equation} 
where $\mu_r$ is the real part of $\mu$. We have a stationary bifurcation for  $\vert \mu \vert =  \vert \nu \vert$ if $\mu_r < 0$, a Hopf bifurcation for $\mu_r = 0$ if $\vert \mu \vert > \vert \nu \vert > 0$,  and a codimension-two bifurcation for  $\mu_r = 0$ and $\mu_i^2 = \vert \nu \vert^2$. Writing $A = R \exp i\phi$, the stability of finite amplitude stationary patterns can be studied in the phase approximation provided the amplitude $R$ is slaved to the phase. The imaginary part of the linear part of (\ref{eqdipquad}) gives
\begin{equation}
\dot{\phi}=\mu_i - \nu_r \sin 2\phi + \nu_i \cos 2\phi.
\end{equation} 
As described in section~\ref{model}, stationary solutions disappear via a saddle-node bifurcation when $\mu_i^2 = \vert \nu \vert^2$ and a limit cycle that corresponds to travelling waves i.e., periodic reversals, is generated. This occurs only if $\mu_i$ increases faster than $\nu_i$ when the reflection symmetry is externally broken. If $\vert \nu_i \vert >  \vert \mu_i  \vert $, the solutions remain stationary. In the context of the dipole-quadrupole interaction in the dynamo problem, a broken symmetry that induces $\vert \nu_i \vert$ much larger than  $\vert \mu_i \vert$ has been found as a mechanism for hemispherical dynamos (Gallet \& P\'etr\'elis 2009). In the case of axisymmetric convection, the broken reflection symmetry is related to the curvature of the rolls.

The above phase approximation breaks down in the vicinity of the codimension-two point. Another type of bifurcation from stationary solutions to a limit cycle takes place in that case (Gambaudo 1985, Guckenheimer \& Homes 1986). 
It is a subcritical Hopf bifurcation that can be easily discriminated from the previous scenario; the limit cycle appears with a finite period whereas the period diverges when it is generated through a saddle-node bifurcation. Stationary and travelling solutions coexist in some parameter range, thus this second scenario displays bistability. Both bifurcation types, saddle-node and subcritical Hopf, have been reported for the axisymmetric convection problem (Siggers 2003) as well as for reversals of the magnetic field in the VKS experiment (Berhanu \textit{et al.} 2009). In numerical simulations of mean-field  dynamo models, the occurence of reversals of the magnetic field close to a codimension-two bifurcation point
has been emphasized by Stefani \& Gerbeth (2005).

As mentioned above, drift bifurcations of cellular patterns also occur in a parity invariant system provided the reflection symmetry is spontaneously broken through a secondary bifurcation. This completes the analogy between drifting patterns in laminar flows and reversals of large scale fields in the presence of turbulent fluctuations that can also occur as a result of an external or a spontaneous broken symmetry. 

Turbulent fluctuations thus do not play a major r\^ole in the geometry of reversals, i.e. to determine the trajectory that connects two solutions of opposite polarity in phase space. In all the cases studied here, this connection involves at least two large scale modes. 
Turbulent fluctuations of course affect the time dependence of reversals. They trigger  random reversals below the saddle-node bifurcation to a limit cycle. However, very similar random behaviours can be obtained with deterministic models involving three large scale modes, so that it would not be so easy to determine the source of randomness in a natural system displaying reversals.

\section*{\large References}

\smallskip
\noindent
Ahlers, G., Grossmann, S. and Lohse, Heat transfer and large scale dynamics in turbulent Rayleigh-B\'enard convection, {\it Rev. Mod. Phys.}, 2009, {\bf 81}, 503--537.

\smallskip
\noindent
Baldwin, M. P., Gray, L. J., Dunkerton, T. J., Hamilton, K., Haynes, P. H., Randel, W. J., Holon, J. R., Alexander, M. J., Hironta, I., Horinouchi, T., Jones, D. B. A., Kinnersley, J. S., Marquardt, C., Sato, K. and Takahashi, M., The Quasi-Biennial Oscillation, {\it Reviews of Geophysics}, 2001, {\bf 39}, 179-229.

\smallskip
\noindent
Berhanu, M., Monchaux, R., Fauve, S., Mordant, N., P\'etr\'elis, F., Chiffaudel, A., Daviaud, F., Dubrulle, B., Mari\'e, L., Ravelet, F., Bourgoin, M., Odier, Ph., Pinton, J.-F., Volk, R., Magnetic field reversals in an experimental turbulent dynamo, {\it Europhys. Lett.}, 2007, {\bf 77}, 59001.	

\smallskip
\noindent
Berhanu, M., Monchaux, R., Bourgoin, M., Odier, Ph., Pinton, J.-F., Plihon, N., Volk, R., Fauve, S., Mordant, N., P\'etr\'elis, F., Aumatre, S., Chiffaudel, A., Daviaud, F., Dubrulle, B., Ravelet, F., Bistability between a stationary and an oscillatory dynamo  in a turbulent flow of  liquid sodium,  {\it J. Fluid Mech.}, 2009, {\bf 641}, 217-226.

\smallskip
\noindent
Breuer, M. and Hansen, U., Turbulent convection in the zero Reynolds number limit,  {\it Europhys.Lett.}, 2009, {\bf 86}, 24004.

\smallskip
\noindent
 Brunhes, B., Recherches sur la direction d'aimantation des roches volcaniques. {\it J. de Phys. Theor. App.}, 1906, {\bf 5}, 705-724.

\smallskip
\noindent
Caroli, B., Caroli C. and Fauve, S., On the phenomenology of tilted domains in lamellar eutectic growth, {\it J. Physique I}, 1992, {\bf 2}, 281--290.

\smallskip
\noindent
Chandrasekhar, S.,  {\it Hydrodynamic and hydromagnetic stability}, 1961, (Clarendon Press, Oxford).

\smallskip
\noindent
Coullet, P., Goldstein, R. E.  and Gunaratne, G. H., Parity-breaking transitions of modulated patterns in hydrodynamic systems, {\it Phys. Rev. Lett.}, 1989, {\bf 63}, 1954--1957.

\smallskip
\noindent
Dormy, E., Valet J.-P. \& Courtillot V.,   Numerical models of the geodynamo and observational constraints.  {\it Geochem. Geophys. Geosyst.}, 2000, \textbf{1}, 2 2000GC000062.

\smallskip
\noindent
Douady, S., Fauve, S. and Thual, O.,  Oscillatory phase modulation of parametrically forced surface waves, {\it Europhys. Lett.}, 1989, {\bf 10}, 309--315.


\smallskip
\noindent
Fauve, S., Laroche, C. and Libchaber, A. and Perrin, B., Chaotic phases and magnetic order in a convective fluid, {\it Phys. Rev. Letters}, 1984, {\bf 52}, 1774-1777.

\smallskip
\noindent
Fauve, S., Douady, S. and Thual, O., Drift instabilities of cellular patterns, {\it J. Physique II}, 1991, {\bf 1}, 311-322.


\smallskip
\noindent
Gallet, B. and P\'etr\'elis, F.,  From reversing to hemispherical dynamos, {\it Phys. Rev. E}, 2009, {\bf 80}, 035302.

\smallskip
\noindent
Gailitis, A., Lielausis, O., Platacis, E.,  Dement'ev, S.,
Cifersons, A., Gerbeth, G., Gundrum, T., Stefani, F., Christen M. and  
Will, G., Magnetic field saturation in the Riga dynamo experiment, 
{\it Phys. Rev. Lett.}, 2001, {\bf 86}, 3024-3027.

\smallskip
\noindent
Gambaudo, J. M., Perturbation of a Hopf bifurcation by an external time-periodic forcing, {\it J. Diff. Eqs.}, 1985, {\bf 57}, 172-199.

\smallskip
\noindent
Ghil, M. \& Childress, S.,  \textit{Topics in geophysical fluid dynamics: atmospheric dynamics, dynamo theory, and climate dynamics}. Appl. Math. Sci. vol. 60, 1987 (New York: Springer Verlag).

\smallskip
\noindent
Gissinger C., Dormy E., Fauve S., Morphology of field reversals in turbulent dynamos, {\it Europhys. Lett.}, 2010, {\bf  90}, 49001.  

%

\smallskip
\noindent
Glatzmaier, G.A. and Roberts, P.H., 
A three-dimensional self-consistent computer simulation of a geomagnetic field reversal,
{\it Nature}, 1995, {\bf 377}, 203-209.

\smallskip
\noindent
Grebogi, C., Ott, E. and Yorke, J. A., Chaotic attractors in crisis, {\it Phys. Rev. Lett.}, 1982, {\bf 48}, 1507.

\smallskip
\noindent
Guckenheimer, J. and Holmes, P.,   {\it Nonlinear Oscillations, Dynamical Systems and Bifurcations of Vector Fields}, 1986 (Springer-Verlag, New-York).

\smallskip
\noindent
Hughes D. W. and Proctor, M.R.E., A low-order model of the shear instability of convection: chaos and the effect of noise, {\it Nonlinearity}, 1990, {\bf 3}, 127-153.

\smallskip
\noindent
Krishnamurti, R. and Howard, L. N.,
Large-scale flow generation in turbulent convection.{\it Proc. Natl. Sci. USA}, 1981, {\bf 78},1981--1985.

\smallskip
\noindent
Labb\'e, R., Pinton, J. F.  and Fauve, S., Study of the von Karman flow between coaxial corotating disks,  {\it Phys. Fluids}, 1996,  {\bf 8}914-922. 

\smallskip
\noindent
Larmor, J., How could a rotating body such as the
sun become a magnet?, Rep. $87^{\rm th}$ Meeting Brit. Assoc. Adv.
Sci., Bornemouth, Sept. 9-13, 1919, pp. 159-160, 1919 (London: John Murray).

\smallskip
\noindent
Li, J., Sato, T. and Kageyama, A.,
Repeated and sudden reversals of the dipole field generated by a spherical dynamo action,   {\it Science}, 2002, {\bf 295}, 1887-1890.

\smallskip
\noindent
Liu, B. and Zhang, J., Self-induced cyclic reorganization of free bodies through thermal convection, {\it Phys. Rev. Lett.}, 2008, {\bf 100}, 244501.


\smallskip
\noindent
Lorenz, E., Deterministic non periodic flow,  {\it Journal of the Atmospheric Sciences}, 1963,  {\bf 20}, 130-141. 

\smallskip
\noindent
Mishra, P. K., De, A. K., Verma, M. K. and Eswaran, V., Dynamics of reorientations and reversals of large-scale flow in Rayleigh-B\'enard convection, {\it J. Fluid Mech.}, 2010, (in press).

\smallskip
\noindent
Moffatt, H. K., {\it Magnetic Field Generation in Electrically Conducting Fluids}, 1978 (Cambridge University Press).

\smallskip
\noindent
Monchaux, R., Berhanu, M.,  Bourgoin, M.,  Moulin, M.,  Odier, Ph.,  Pinton, J.-F., Volk, R., Fauve, S., Mordant, N., P\'etr\'elis, F.,   Chiffaudel, A., Daviaud, F., Dubrulle, B., Gasquet, C. and Mari\'e,  L., Generation of magnetic field by dynamo action in a turbulent flow of liquid sodium, {\it Phys. Rev. Lett.}, 2007, {\bf 98}, 044502.


\smallskip
\noindent
Mutabazi, I., Hegseth, J. J. and Andereck, C. D., Pattern formation in the flow between two horizontal coaxial cylinders with a partially filled gap,
{\it Phys. Rev. A}, 1988, {\bf 38}, 4752--4760.

\smallskip
\noindent
Nishikawa, N. and  Kusano, K.,
Simulation study of the symmetry-breaking instability and the dipole field reversal in a rotating spherical shell dynamo,
{\it Physics of Plasma}, 2008,  {\bf 15}, 082903. 


\smallskip
\noindent
Nozi\`eres, P., Reversals of the Earth's magnetic field: an attempt at a relaxation model, 
{\it Physics of the Earth and Planetary Interior},  1978,  {\bf 17}, 55--74. 


\smallskip
\noindent
Paul, S, Kumar, K., Verma, M. K., Carati, D., De, A. K. and Eswaran, V., Chaotic travelling rolls in Rayleigh-B\'enard convection, {\it Pramana J. Phys.}, 2010, {\bf 74}, 75-82.

\smallskip
\noindent
P\'etr\'elis F. and Fauve S.,  Chaotic dynamics of the magnetic field generated by dynamo action in a turbulent flow, {\it J. Phys.: Condens. Matter}, 2008,  {\bf 20}  494203.

\smallskip
\noindent
P\'etr\'elis, F. , Dormy, E., Valet, J.-P. \& Fauve, S., Simple mechanism for the reversals of Earth magnetic field, 2009, {\it Phys. Rev. Lett.} {\textbf 102}, 144503.

\smallskip
\noindent
P\'etr\'elis F. and Fauve S., Mechanisms for magnetic field reversals, {\it Philosophical Transactions of the Royal society A}, 2010,   {\textbf 368}, 1595-1605.

\smallskip
\noindent
Rabaud, M, Michalland, S. and Couder, Y.,  Dynamical regimes of directional viscous fingering: Spatiotemporal chaos and wave propagation,
{\it Phys. Rev. Lett.}, 1990, {\bf 64}, 184--187.

\smallskip
\noindent
Ravelet, F., Mari\'e, L., Chiffaudel, A. and Daviaud, F., Multistability and Memory Effect in a Highly Turbulent Flow: Experimental Evidence for a Global Bifurcation, {\it Phys. Rev. Lett.}, 2004, {\bf 93}, 164501.

\smallskip
\noindent
Ravelet, F.,  Berhanu, M., Monchaux, R., Aumaitre S., Chiffaudel, A., Daviaud, F, Dubrulle, B., Bourgoin, M., Odier, Ph., Plihon, N., Pinton, J.-F, Volk, R., Fauve, S., Mordant, N. and Petrelis F.,  Chaotic dynamos  generated by a turbulent flow of liquid sodium, {\it Phys. Rev. Lett.}, 2008, {\bf 101}, 074502.

\smallskip
\noindent
Rikitake, T., Oscillations of a system of disc dynamos, {\it Proc. Camb. Phil. Soc.}, 1958,  {\bf 54}, 89-105.

\smallskip
\noindent
Sarson, G.R. and Jones, C.A., 
A convection driven geodynamo reversal model, 
{\it Physics of the Earth and Planetary Interior}, 1999, {\bf 111}, 3-20.

\smallskip
\noindent
Siggers, J. H.,  Dynamics of target patterns in low-Prandtl-number convection,  {\it J. Fluid Mech.},  2003,  {\bf 475}, 357-375.

\smallskip
\noindent
Simon, A. J.,  Bechhoefer, J. and Libchaber, A., Solitary Modes and the Eckhaus Instability in Directional Solidification, {\it Phys. Rev. Lett.}, 1988, {\bf 61}, 2574-2577.

\smallskip
\noindent
Sommeria J.,  Experimental study of the two-dimensional inverse energy cascade in a squarre box,  {\it J. Fluid Mech.}, 1986, {\bf 170}, 139-168.

\smallskip
\noindent
Stefani F. and Gerbeth G., Asymmetric Polarity Reversals, Bimodal Field Distribution, and Coherence Resonance
in a Spherically Symmetric Mean-Field Dynamo Model, {\it Phys. Rev. Lett.}, 2005, {\bf 94}, 184506.


\smallskip
\noindent
Stieglitz, R. and M\"uller,  U., Experimental demonstration of a  homogeneous two-scale dynamo, {\it Phys. Fluids}, 200, {\bf 13}, 561-564.


\smallskip
\noindent
Sugiyama, K., Ni, R.,  Stevens, R. J. A. M., Chan, T. S., Zhou, S.-Q.,  Xi, H.-D., Sun, C., Grossmann, S.,  Xia, K. Q. and Lohse, D., Flow Reversals in Thermally Driven Turbulence, {\it Phys. Rev. Lett.}, 2010,  {\bf 105}, 034503.

\smallskip
\noindent
Tabeling P., Perrin B., Fauve S., Instability of a Linear Array of Forced Vortices, {\it Europhys. Lett.}, 1987, {\bf 3} (4),  459-465.

\smallskip
\noindent
Tritton, D. J. {\it Physical Fluid Dynamics} 1977 (van Nostrand Reinhold Company, New York 1977). 

\smallskip
\noindent
Tuckerman, L. S.  and  Barkley, D., Global bifurcation to traveling waves in axisymmetric convection,  {\it Phys. Rev. Lett.}, 1988, {\bf 61}, 408-411.

\smallskip
\noindent
Valet, J.-P., Meynadier, L. \& Guyodo, Y. 2005 Geomagnetic field strength and reversal rate over the past 2 Million years, {\it Nature}  {\bf 435}, 802-805.

\smallskip
\noindent
Vallis, G. K., El Nino: A Chaotic Dynamical System?
\textit{Science},  1986, \textbf{232}, 243--245.

\smallskip
\noindent
Wicht, J. and Olson, P.,  A detailed study of the polarity reversal mechanism in a numerical dynamo model, {\it Geochemistry, Geophysics, Geosystems}, 2004,  {\bf 5}, Q03H10.

\end{document}